\documentclass[preprint,prb,amsmath,amssymb]{revtex4}

\usepackage{graphicx}

\begin{document}

\title{Asymptotic solution of the diffusion equation in slender impermeable 
tubes of revolution. I. The leading-term approximations}

\author{Sergey~D.~Traytak} \thanks{Electronic mail: sergtray@mail.ru}

\affiliation{Centre de Biophysique Mol\'eculaire, CNRS-UPR4301,
Rue C. Sadron, 45071, Orl\'eans, France}
\affiliation{Le STUDIUM (Loire Valley Institute for Advanced Studies), 3D av. de la Recherche scientifique, 45071, Orl\'eans, France}
\affiliation{Semenov Institute of Chemical Physics RAS,
4 Kosygina St.,
117977 Moscow, Russia}

\begin{abstract}
The anisotropic 3D equation describing the pointlike particles diffusion in slender impermeable tubes of revolution with cross section smoothly depending on the longitudial coordinate is the object of our study. We use singular perturbations approach to find the rigorous asymptotic expression for the local particles concentration as an expansion in the ratio of the characteristic transversal and longitudial diffusion relaxation times. The corresponding leading-term approximation is a generalization of well-known Fick-Jacobs approximation. This result allowed us to delineate the conditions on temporal and spatial scales under which the Fick-Jacobs approximation is valid. A striking analogy between solution of our problem and the method of inner-outer expansions for low Knudsen numbers gas kinetic theory is established. With the aid of this analogy we clarify the physical and mathematical meaning of the obtained results.
\end{abstract}

\maketitle

\section{Introduction}

The problem of approximate reduction of the time-dependent 3D equation
describing the local concentration field $C\left( \mathbf{x},t\right) $ of
pointlike particles diffusing in a tube of varying with the longitudinal
coordinate $z$ cross section to an effective time-dependent 1D equation
appeared to be fairly tricky. For the first time, following the main idea of
Fick's approach, such kind of 1D equation was derived in 1935 by Jacobs.\cite%
{Jacobs:1967} Particularly for a channel with a shape of a surface of
revolution the relevant 1D equation reads%
\begin{equation}
\frac{\partial c\left( z,t\right) }{\partial t}=\frac{\partial }{\partial z}%
D\left\{ A\left( z\right) \frac{\partial }{\partial z}\left[ \frac{c\left(
z,t\right) }{A\left( z\right) }\right] \right\} ,  \label{in1}
\end{equation}%
where $D$ is the translational diffusion coefficient in space with no
constraints, $A\left( z\right) =\pi \left[ r\left( z\right) \right] ^{2}$ is
the area of the tube cross-section $\Sigma _{z}$ of radius $r\left( z\right) 
$ at a given point $z$ of the symmetry axis. The corresponding reduced
concentration $c\left( z,t\right) $ is calculated by the formula 
\begin{equation}
c\left( z,t\right) =\int\limits_{\Sigma _{z}}C\left( x,y,z,t\right) dxdy.
\label{in2}
\end{equation}%
Nowadays the reduced diffusion equation (\ref{in1}) is commonly referred to
as the Fick-Jacobs equation (FJE). In addition we will call Eq. (\ref{in1})
a classical form of the FJE. It is interesting that, if we do not take into
account a few works on this subject, for decades the FJE remained almost
unclaimed. Situation has been changed drastically after 1992 when well-known
Zwanzig's article renewed the problem and stimulated considerable interest
to this topic.\cite{Zwanzig:1992} Soon it turned out that the problem on
diffusion in a tube of varying cross section is of great importance for
numerous applications dealing with artificial and natural transport
processes and thence many researchers studied it within different facets of
theory and applications.\cite%
{Grathwohl:1998,Konkoli:2005,Berezhkovskii:2007,Dendrites:2008,Burada:2008,Ai:2008,Wang:2009,Mondal:2010,Dagdug:2010,Biess:2011,Dagdug:2011,Berezhkovskii:2012}
Even now these kind of investigations are close to the top among hot
research topics on the diffusion-influenced processes in confined regions.%
\cite{Barzykin:2013}

In his seminal paper Zwanzig drew attention to the fact that Jacobs
derivation is rather heuristic and, besides, it is completely free of
impermeable wall boundary condition that should be imposed on the solution
of the original higher dimensional diffusion equation. Moreover, Jacobs did
not present any reasons for choosing the center line of the tube.\cite%
{Zwanzig:1992} Taking into account that the FJE has exactly the same
mathematical structure as the Smoluchowski equation for diffusion in a 1D
potential field, Zwanzig derived the FJE starting from the general diffusion
equation with a potential.\cite{Zwanzig:1992} According to Zwanzig the FJE
may be presented in the form%
\begin{equation}
\frac{\partial c\left( z,t\right) }{\partial t}=\frac{\partial }{\partial z}%
D\left\{ e^{-\frac{U\left( z\right) }{k_{B}T}}\frac{\partial }{\partial z}%
\left[ e^{\frac{U\left( z\right) }{k_{B}T}}c\left( z,t\right) \right]
\right\} .  \label{in2a}
\end{equation}%
In (\ref{in2a}) $U\left( z\right) $ is so-called entropy potential defined as%
\[
U\left( z\right) =-k_{B}T\ln A\left( z\right) , 
\]%
where $k_{B}$ and $T$ are the Boltzmann constant and the absolute
temperature. Assuming that the channel radius varies slowly with increasing
of the longitudinal variable $z$ , i.e., 
\begin{equation}
\left\vert r^{^{\prime }}\left( z\right) \right\vert \ll 1,  \label{in2b}
\end{equation}%
hereafter in the paper $\psi ^{^{\prime }}\left( \varsigma \right) :=d\psi
\left( \varsigma \right) /d\varsigma $, he also proposed a generalized form
of the FJE. For particular case of 3D tube the original diffusion constant $%
D $ in the FJE (\ref{in1}) was replaced by a spatially dependent effective
diffusion coefficient 
\[
D_{Zw}\left( z\right) \approx \frac{D}{1+\frac{1}{2}\left[ r^{^{\prime
}}\left( z\right) \right] ^{2}}. 
\]%
Later, in 2001 Reguera and Rubi developed this idea using\ some
nonequilibrium thermodynamics reasons and obtained the corrected FJE with 
\[
D_{R-R}\left( z\right) \approx \frac{D}{\sqrt{1+\left[ r^{^{\prime }}\left(
z\right) \right] ^{2}}} 
\]%
in case of 3D symmetric tubes.\cite{Reguera:2001}

Further corrections to the FJE were obtained by Kalnay and Percus with the
help of so-called mapping technique, which differs from Zwanzig's entropy
barrier theory.\cite{Kalnay:2005,Kalnay2:2005,Kalnay:2006} This mapping
procedure has been performed for the anisotropic diffusion equation without
a potential%
\begin{equation}
\frac{\partial C}{\partial t}=\nabla \cdot \left( \mathbf{D\cdot \nabla }%
C\right) ,  \label{in3}
\end{equation}%
where $\mathbf{D}$ is the translational diffusion tensor matrix. As in Refs.
17 and 19 we suppose here that the diffusion matrix is diagonal with
transverse isotropy, i.e., $\mathbf{D}=$diag$\left( D_{x},D_{y},D_{z}\right) 
$, and that $D_{x}=D_{y}=D_{\perp }$ is the transverse and $%
D_{z}=D_{\parallel }$ is the longitudinal translational diffusion
coefficient, respectively. In their study Kalnay and Percus assumed also
that 
\begin{equation}
\varepsilon =D_{\parallel }/D_{\perp }\ll 1,  \label{in4}
\end{equation}%
which allowed them to suggest that the transverse concentration profile
equilibrates quickly and so-called Zwanzig's factorization\cite{Zwanzig:1992}
holds true. Moreover, this is a quasi steady-state theory, i.e., field $%
C\left( \mathbf{x},t\right) $ is assumed to be an explicitly
time-independent. Time dependence is presented in this function implicitly
by functional of the reduced concentration $c\left( z,t\right) $ only. Under
given assumptions in the limit $\varepsilon \rightarrow 0$ diffusion
equation (\ref{in3}) with reftecting wall condition is reduced to the
corresponding FJE. In its turn for higher-order terms in $\varepsilon $
Kalnay and Percus derived a generalized 1D equation that contains all higher
derivatives with respect to $z$ of the tube radius $r\left( z\right) $ and
reduced concentration $c\left( z,t\right) $.\cite%
{Kalnay:2005,Kalnay2:2005,Kalnay:2006} Kalnay and Percus drew attention to
the fact that the problem "requires an analysis of the short-time behavior",
but they did not deal with this question.\cite{Kalnay:2005} The projection
method has been used by Dagdug and Pineda to find more general effective
diffusion coefficient for the FJE, describing the unbiased motion of
pointlike particles in 2D slender tilted asymmetric channels of varying
width formed by straight wall.\cite{Dagdug1:2012} In the subsequent paper of
the same authors, to test the validity of obtained formulae, a comparison of
these analytical results against Brownian dynamics simulation results were
performed.\cite{Dagdug:2012}

The biased diffusion transport of pointlike particles under the influence of
a constant and uniform force field in 2D and 3D narrow spatially periodic
channels of varying cross section is also rather well investigated.\cite%
{Rubi:2007,Martens:2011,Martens2:2011}

It is worth noting that on account of a mathematical difficulties in solving
the original problem for arbitrary tube radius $r\left( z\right) $ an
effective 1D description for the simplified case when the tubes composed of
some number of contacting equal spheres\cite{Berezhkovskii:2008} or
cylindrical sections of different diameters was investigated.\cite%
{Barzykin:2009,Makhnovskii:2010} Further generalization of the previous
research to the case of a periodically expanded conical tube was recently
reported.\cite{Barzykin:2013} It is important that tubes of this shape may
be utilized as a controlled drug release device.\cite{Barzykin:2013} The
interested reader can find numerous references to the previous analytical
and numerical studies in a recent paper by Kalnay.\cite{Kalnay:2013}

The analysis of the literature showed that, despite the great amount of
publications devoted to the topic, rigorous mathematical study of the
corresponding boundary value problem for all spatial and temporal scales is
still missing. Thus, the purpose of this paper is twofold. Firstly using
rigorous technique of the matched asymptotic expansions \cite%
{Lagerstrom:1988,Ilin:1992} we consider the 3D anisotropic diffusion
equation (\ref{in3}) which describes the diffusion of pointlike particles
into a tube with impermeable wall having the shape of a surface of
revolution. Secondly accurate asymptotic solution of the original 3D problem
for all spatial and temporal scales allows us to elucidate the role,
physical and mathematical sense and lastly accuracy of the leading-term
approximation and, particularly, the validity of the Fick-Jacobs
approximation.

The paper is organized in the following way. Section II contains the full
mathematical statement of the corresponding boundary value problem. In Sec.
III by means of singular perturbations approach we give the detailed
preliminary ideas for asymptotic solution of the posed problem. Section IV
devotes to the asymptotic solution in the outer subdomain and, particularly,
derivation of the Fick-Jacobs equation. In Sec. V and Sec. VI we study
solution in spatial and temporal diffusive boundary layers, respectively, to
derive, in particular, the appropriate boundary and initial conditions for
solution of the Fick-Jacobs equation. Section VII presents determination of
the corner asymptotic solutions. In Sec. VIII the main result of the paper
the leading-term approximation is reported. This section also comprises
criteria for validity of the Fick-Jacobs approximation and establishes a
profound analogy of the problem at issue with the gas kinetic theory for low
Knudsen numbers. Finally the main concluding remarks are made in Sec. IX.
Some subsidiary classical mathematical facts are given in Appendix.

\section{Statement of the problem}

\begin{figure}[t!]
\centering
\resizebox{90mm}{!}{\includegraphics[clip]{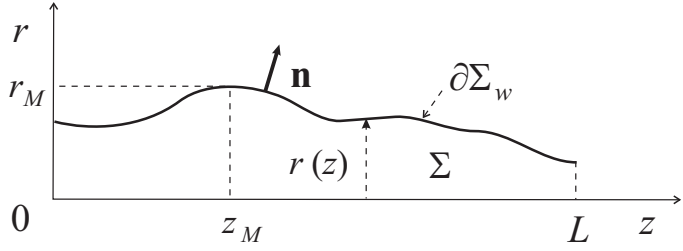}} 
\caption{Geometric sketch of the problem.}
\label{fig:geom}
\end{figure}

Consider the pointlike particles diffusion in a 3D tube of length $L$, which
wall is obtained by rotation of the line 
\begin{equation}
r\left( z\right) =r_{M}R\left( z/L\right)  \label{sp1}
\end{equation}%
around the $z$ axis (see Fig.~\ref{fig:geom}). We assume function $r\left( z\right) $ to
be smooth enough and introduced the maximum value of this function 
\[
r_{M}:=r\left( z_{M}\right) =\max_{z\in \left[ 0,L\right] }r\left( z\right) 
\]%
which fully characterizes the transverse size of the tube. It is evident
that 
\[
0<R\left( z/L\right) \leq 1=R\left( z_{M}/L\right) 
\]%
for all $z\in \left[ 0,L\right] $. By definition we shall call tube slender
(narrow) when 
\[
r_{M}\ll L. 
\]%
In the cylindrical coordinate system $\left( r,\phi ,z\right) $ connected
with the $z$ axis the tube region is 
\[
\Sigma :=\left\{ 0<r<r\left( z\right) \right\} \times \left( 0<z<L\right)
\times \left( 0<\phi <2\pi \right) . 
\]
A cross section of the tube at any fixed value $z$ is $\Sigma _{z}:=\left\{
0<r<r\left( z\right) \right\} \times \left( 0<\phi <2\pi \right) $ and $%
\partial \Sigma _{w}:=\left\{ r=r\left( z\right) ,\phi \in \left( 0,2\pi
\right) ,z\in \left[ 0,L\right] \right\} $ is the tube wall. Moreover we
suppose that the local concentration of diffusing particles $C\left( \mathbf{%
x},t\right) $ possesses the axial symmetry and therefore actually we shall
treat here the 2D time-dependent diffusion equation.

Thus the anisotropic diffusion equation (\ref{in3}), for the chosen
cylindrical coordinates reads 
\begin{equation}
\frac{\partial C}{\partial t}=D_{\perp }\frac{1}{r}\frac{\partial }{\partial
r}\left( r\frac{\partial C}{\partial r}\right) +D_{\parallel }\frac{\partial
^{2}C}{\partial z^{2}}  \label{sp2}
\end{equation}%
in the space-time domain $\Sigma _{t}:=\Sigma \times \left( t>0\right) $. It
is clear that at the $z$ axis one should take into consideration conditions
of regularity and axial symmetry of solution, respectively 
\begin{equation}
\lim_{r\rightarrow 0}C<\infty ,\qquad \lim_{r\rightarrow 0}\frac{\partial C}{%
\partial r}=0.  \label{sp3}
\end{equation}

On the wall of the tube $\partial \Sigma _{w}$ we impose the common
reflecting boundary condition 
\begin{equation}
\left. \left( \mathbf{n\cdot j}\right) \right\vert _{\partial \Sigma _{w}}=0,
\label{sp4}
\end{equation}%
where $\mathbf{n}$ being the outer-pointing unit normal with respect to $%
\partial \Sigma _{w}$ (see Fig.~\ref{fig:geom}) and $\mathbf{j=-D\cdot }\nabla C$ is the
local diffusing flux of particles. One can see that the tube wall $\partial
\Sigma _{w}$ may be defined analytically as 
\[
w\left( r,z\right) =r-r_{M}R\left( z/L\right) =0. 
\]%
It is well known that for all nonsingular points $\left( r,z\right) \in $ $%
\partial \Sigma _{w}$ ($\nabla w\left( r,z\right) \neq 0$) the unit normal
may be calculated as $\mathbf{n=}\nabla w/\left\Vert \mathbf{\nabla }%
w\right\Vert $. Taking this into account we can rewrite condition (\ref{sp4}%
) as the orthogonality condition in cylindrical coordinates 
\begin{equation}
\left. \left[ \frac{\partial C}{\partial r}-r_{M}\frac{D_{\parallel }}{%
D_{\perp }}\frac{\partial C}{\partial z}\frac{d}{dz}R\left( z/L\right) %
\right] \right\vert _{w=0}=0.  \label{sp4a}
\end{equation}

To complete the problem statement one has to impose the initial condition 
\begin{equation}
\left. C\right\vert _{t=0}=C_{0}\left( r,z\right) \qquad \text{ in }\Sigma
\label{sp5}
\end{equation}%
and, for definiteness, Diriclet boundary conditions on the ends of the tube 
\begin{equation}
\left. C\right\vert _{z=0}=C_{1}\left( r,t\right) ,\quad \left. C\right\vert
_{z=L}=C_{2}\left( r,t\right) .  \label{sp6}
\end{equation}%
We assume that all given functions $C_{0}\left( r,z\right) $, $C_{1}\left(
r,t\right) $ and $C_{2}\left( r,t\right) $ are continuous in their domains
of definition. Hence according to the maximum principle for the diffusion
equation we see that $C\in \left[ C_{m},C_{M}\right] $, where $%
C_{m}:=\min_{\partial \Sigma _{t}}\left\{ C_{0}\left( r,z\right)
,C_{1}\left( r,t\right) ,C_{2}\left( r,t\right) \right\} $ and $%
C_{M}:=\max_{\partial \Sigma _{t}}\left\{ C_{0}\left( r,z\right)
,C_{1}\left( r,t\right) ,C_{2}\left( r,t\right) \right\} $.

One can see that, due to complex geometry of the tube wall, analytical
solution of the posed boundary value problem (\ref{sp2})-(\ref{sp6}) is not
feasible in general case. That is why the FJE corresponding to Eq. (\ref{sp2}%
)

\begin{equation}
\frac{\partial c\left( z,t\right) }{\partial t}=D_{\parallel }\frac{\partial 
}{\partial z}\left\{ A\left( z\right) \frac{\partial }{\partial z}\left[ 
\frac{c\left( z,t\right) }{A\left( z\right) }\right] \right\}  \label{sp7}
\end{equation}%
plays an important role in applications. Note that in a special case of
round cylindrical channel the FJE (\ref{sp7}) simplifies to the common 1D
second Fick's equation 
\begin{equation}
\frac{\partial c\left( z,t\right) }{\partial t}=D_{\parallel }\frac{\partial
^{2}c\left( z,t\right) }{\partial z^{2}}.  \label{sp9}
\end{equation}

Our main objective with this paper is to construct a rigorous iterative
procedure for asymptotic solution of the problem (\ref{sp2})-(\ref{sp6}) in
case of a slender tube. Particularly this solution entails straightforwardly
the FJE (\ref{sp7}) with appropriate initial and boundary conditions.
Besides, we will find criteria for validity of the corresponding
approximation $c\left( z,t\right) $.

\section{Formulation as a singular perturbed problem}

\subsection{Non-dimensionalization of the problem}

In order to perform the asymptotic solution of the posed boundary value
problem (\ref{sp2})-(\ref{sp6}), we need to nondimensionalize it. It is
expedient to rewrite this problem for dimensionless variables using the
following scales 
\[
\rho =r/r_{M},\qquad \xi =z/L,\qquad \tau =t/t_{L}, 
\]%
where $t_{L}=L^{2}/D_{\parallel }$ is the characteristic longitudinal time
for the diffusion length $L$. Moreover it is also convenient to treat the
normalized dimensionless local concentration 
\[
u\left( \rho .\xi .\tau \right) =C\left( \rho .\xi .\tau \right) /C_{M}. 
\]%
Accordingly, Eq. (\ref{sp2}) takes the form 
\begin{equation}
\left( \mathcal{L}_{\rho }+\epsilon \mathcal{L}_{F}\right) u=0,\quad \text{
in }\Sigma _{\tau },  \label{nd4}
\end{equation}%
where the unperturbed operator 
\begin{equation}
\mathcal{L}_{\rho }:=-\frac{1}{\rho }\frac{\partial }{\partial \rho }\left(
\rho \frac{\partial }{\partial \rho }\right)  \label{nd5}
\end{equation}%
is the radial part of the 2D Laplacian in polar coordinates. For the
notation convenience in Eq. (\ref{nd4}) and hereafter we define the
dimensionless 1D Fick operator 
\[
\mathcal{L}_{F}:=\frac{\partial }{\partial \tau }-\frac{\partial ^{2}}{%
\partial \xi ^{2}}. 
\]%
It is evident that for this problem $\epsilon \mathcal{L}_{F}$ being the
perturbation operator. We shall show below that as $\epsilon \rightarrow 0$
unperturbed operator $\mathcal{L}_{\rho }$ and perturbation $\epsilon 
\mathcal{L}_{F}$ determine the fast and slow behavior of the desired
solution $u\left( \rho .\xi .\tau \right) $, respectively.

Initial and boundary conditions now read 
\begin{equation}
\left. u\right\vert _{\tau =0}=g_{0}\left( \rho ,\xi \right) ,  \label{nd6}
\end{equation}%
\begin{equation}
\left. u\right\vert _{\xi =0}=g_{1}\left( \rho ,\tau \right) ,\qquad \left.
u\right\vert _{\xi =1}=g_{2}\left( \rho ,\tau \right) ,  \label{nd7}
\end{equation}%
\begin{equation}
\left. \left[ \frac{\partial u}{\partial \rho }-\epsilon \frac{\partial u}{%
\partial \xi }R^{^{\prime }}\left( \xi \right) \right] \right\vert _{\rho
=R\left( \xi \right) }=0,  \label{nd8}
\end{equation}%
\begin{equation}
\left. u\right\vert _{\rho =0}<\infty ,\qquad \left. \frac{\partial u}{%
\partial \rho }\right\vert _{\rho =0}=0.  \label{nd9}
\end{equation}%
Hereafter we denote $g_{\gamma }=C_{\gamma }/C_{M}$ ($\gamma =0,1,2$).

In Eq. (\ref{nd4}) and boundary condition (\ref{nd8}) we introduced a new
dimensionless parameter $\epsilon $ describing well the system under study 
\begin{equation}
\epsilon =\epsilon _{s}^{2}\frac{D_{\parallel }}{D_{\perp }},  \label{nd10}
\end{equation}%
where 
\begin{equation}
\epsilon _{s}=\frac{r_{M}}{L}\ll 1  \label{nd10a}
\end{equation}%
is the slenderness ratio (the relative thickness of the tube) for a narrow
tube.\cite{Wang:2009} Note that, to make the Zwanzig's factorization more
plausible, it is usually assumed that relationship (\ref{in4}) holds true.%
\cite{Kalnay:2005,Kalnay2:2005,Kalnay:2006} However it often happens in
applications that $D_{\parallel }/D_{\perp }>1$.\cite{Bihan:2003} In any
case we consider here the slenderness ratio $\epsilon _{s}$ is small enough
to make $\epsilon $ small. Furthermore there is another important physical
meaning of the introduced parameter $\epsilon $. To clarify this we rewrite (%
\ref{nd10}) as 
\begin{equation}
\epsilon =\frac{t_{tr}}{t_{L}}\ll 1,  \label{nd11}
\end{equation}%
where $t_{tr}=r_{M}^{2}/D_{\perp }$ is the characteristic transversal time
for the diffusion length $r_{M}$. Physically inequality (\ref{nd11}) means
that the diffusive relaxation along the transversal direction occurs much
faster than that along axis $z$. In this connection we can call $\epsilon $
a relaxation parameter.

Simple inspection shows that for $\epsilon \rightarrow 0$ the posed problem (%
\ref{nd4})-(\ref{nd9}) is a singularly perturbed one.\cite%
{Lagerstrom:1988,Ilin:1992} Really, if we just set $\epsilon =0$ in (\ref%
{nd4}) we obtain unperturbed equation with a general solution, which cannot
satisfy nor initial nor boundary conditions (\ref{nd6})-(\ref{nd8}). To
study this problem we shall apply method of matched asymptotic expansions,
which proved to be a powerful tool for solution of many singularly perturbed
problems concerning the diffusion-influenced processes.\cite%
{Traytak:1990,TraytakCPL:1991,Traytak:2004,TraytakCM:2007}

\subsection{Subdomains for the asymptotic solution. Diffusion boundary layers%
}

Taking into account singularity of the perturbed problem (\ref{nd4})-(\ref%
{nd9}) one can see that the diffusion equation exhibits certain diffusion
boundary layers, i.e., subdomains of rapid change in the solution and its
derivatives. The location and thickness of the boundary layer depends on a
small parameter inherent in the problem under consideration (in our case
this is $\epsilon $). For instance, it follows immediately from the general
form of perturbed operator and conditions (\ref{nd6}) and (\ref{nd7}) that
our problem possesses one temporal and two spatial boundary layers.\cite%
{Ilin:1992}

To facilitate an understanding of the further study we shall use another
decomposition of the space-time domain $\Sigma _{\tau }=\left\{ 0<\rho
<R\left( \xi \right) \right\} \times \Omega $, where the semi-infinite strip 
$\Omega :=\left( 0<\xi <1\right) \times \left( \tau >0\right) $ is its 2D
cross section. For the boundary of the domain $\Omega $ we have $\partial
\Omega =\partial \Omega _{\tau }\cup \partial \Omega _{\xi }$, where 
\[
\partial \Omega _{\tau }:=\left\{ \left( \xi ,\tau \right) :\tau =0\right\} 
\]%
is the temporal boundary 
\[
\partial \Omega _{\xi }=\partial \Omega _{0}\cup \partial \Omega _{1} 
\]%
is the spatial boundary comprising two connected components at the endpoints
of the tube 
\[
\partial \Omega _{0}:=\left\{ \left( \xi ,\tau \right) :\xi =0\right\}
,\quad \partial \Omega _{1}:=\left\{ \left( \xi ,\tau \right) :\xi
=1\right\} . 
\]

\begin{figure}[t!]
\centering
\resizebox{90mm}{!}{\includegraphics[clip]{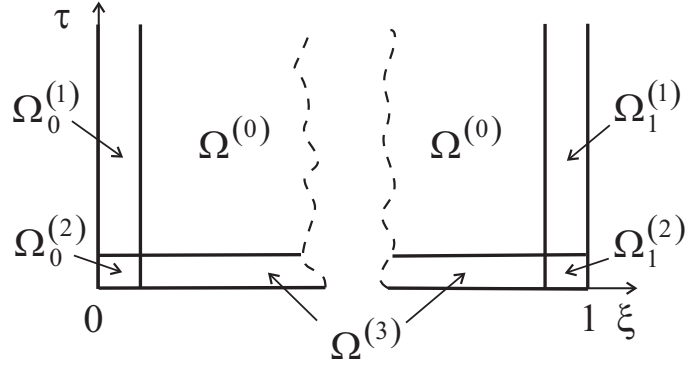}} 
\caption{Depiction of the relationship of the boundary layers to the outer subdomain $\Omega ^{\left( 0\right) }$: spatial boundary layers $\Omega _0^{\left( 1\right) }$, $\Omega _1^{\left( 1\right) }$; corner boundary layers $\Omega _0^{\left( 2\right) }$, $\Omega _1^{\left( 2\right) }$ and temporal boundary layer $\Omega _0^{\left( 3\right) }$.}
\label{fig:dom}
\end{figure}

Consider the structure of the diffusion boundary layer (see Fig.~\ref{fig:dom}). It is
clear that we can decompose the 2D domain of variables $\left( \xi ,\tau
\right) $ as follows: 
\begin{equation}
\Omega =\Omega ^{\left( 0\right) }\cup \Omega ^{\left( b\right) }.
\label{sd1}
\end{equation}%
Here $\Omega ^{\left( 0\right) }$ is a subdomain, where one does not expect
rapid change in the solution and its derivatives and so relevant solution
depends on slow variables $\left( \xi ,\tau \right) $ only. On the other
hand a subdomain $\Omega ^{\left( b\right) }$ is the diffusion boundary
layer that abutted the boundary\thinspace $\partial \Omega $. Usually in
perturbations theory the boundary layer subdomain $\Omega ^{\left( b\right)
} $ and $\Omega ^{\left( 0\right) }$ are called inner and outer subdomains,
respectively.\cite{Ilin:1992}

Structure of the perturbed equation (\ref{nd4}) allows us to define entirely
the boundary layer 
\begin{equation}
\Omega ^{\left( b\right) }=\Omega _{\xi }^{\left( b\right) }\cup \Omega
_{\tau }^{\left( b\right) },  \label{sd2}
\end{equation}%
where $\Omega _{\tau }^{\left( b\right) }$ is the temporal boundary layer
subdomain abutted the initial values part of the boundary $\partial \Omega
_{\tau }$. In decomposition (\ref{sd2}) the spatial subdomain $\Omega _{\xi
}^{\left( b\right) }$ consists of two strips in the semi-vicinities of the
endpoints $\xi =0$ and $\xi =1$, respectively 
\begin{equation}
\Omega _{\xi }^{\left( b\right) }=\Omega _{0}^{\left( b\right) }\cup \Omega
_{1}^{\left( b\right) }.  \label{sd3}
\end{equation}%
The subsequent partition may be obtained if we introduce the corner
subdomains near vertices $\left( 0,0\right) $ and $\left( 1,0\right) $: $%
\Omega _{0}^{\left( 2\right) }=\Omega _{0}^{\left( b\right) }\cap \Omega
_{\tau }^{\left( b\right) }$ and $\Omega _{1}^{\left( 2\right) }=\Omega
_{1}^{\left( b\right) }\cap \Omega _{\tau }^{\left( b\right) }$, where we
have intersection of spatial and temporal boundary layers. Whence we can
represent a strip near the left endpoint $\xi =0$ as $\Omega _{0}^{\left(
b\right) }=\Omega _{0}^{\left( 1\right) }\cup \Omega _{0}^{\left( 2\right) }$
and similarly a strip near the right endpoint $\xi =1$ as $\Omega
_{1}^{\left( b\right) }=\Omega _{1}^{\left( 1\right) }\cup \Omega
_{1}^{\left( 2\right) }$. Finally for the problem under study we obtain the
following five fold partition of the diffusion boundary layer (see Fig.~\ref{fig:dom}) 
\begin{equation}
\Omega ^{\left( b\right) }=\Omega _{0}^{\left( 1\right) }\cup \Omega
_{0}^{\left( 2\right) }\cup \Omega _{0}^{\left( 3\right) }\cup \Omega
_{1}^{\left( 1\right) }\cup \Omega _{1}^{\left( 2\right) },  \label{sd4}
\end{equation}%
where $\Omega _{0}^{\left( 3\right) }:=\Omega _{\tau }^{\left( b\right)
}\backslash \left( \Omega _{0}^{\left( 2\right) }\cup \Omega _{1}^{\left(
2\right) }\right) $.

Analysis of the posed boundary value problem (\ref{nd4})-(\ref{nd9}) leads
to the following asymptotic definitions of the subdomains at issue.

(1) Outer subdomain for the slow spatial and temporal variables $\xi $ and $%
\tau $%
\[
\Omega ^{\left( 0\right) }:=\left\{ \mathcal{O}\left( \sqrt{\epsilon }%
\right) <\xi ,\mathcal{O}\left( \epsilon \right) <\tau \right\} ; 
\]

(2) Left boundary layer subdomain for the fast spatial variable and for slow
time $\tau $ 
\[
\Omega _{0}^{\left( 1\right) }:=\left\{ \xi <\mathcal{O}\left( \sqrt{%
\epsilon }\right) ,\tau \right\} ; 
\]

(3) Right boundary layer subdomain for the fast spatial variable and for
slow time $\tau $ 
\[
\Omega _{1}^{\left( 1\right) }:=\left\{ 1-\xi <\mathcal{O}\left( \sqrt{%
\epsilon }\right) ,\tau \right\} ; 
\]

(4) Left corner boundary layer subdomain for the fast spatial and temporal
variables 
\[
\Omega _0^{\left( 2\right) }:=\left\{ \xi <\mathcal{O}\left( \sqrt{\epsilon }%
\right) ,\tau <\mathcal{O}\left( \epsilon \right) \right\} ; 
\]

(5) Right corner boundary layer subdomain for the fast spatial and temporal
variables 
\[
\Omega _1^{\left( 2\right) }:=\left\{ 1-\xi <\mathcal{O}\left( \sqrt{%
\epsilon }\right) ,\tau <\mathcal{O}\left( \epsilon \right) \right\} ; 
\]

(6) Initial boundary layer: Low inner subdomain for the fast temporal
variable and slow spatial coordinate $\xi $ 
\[
\Omega _{0}^{\left( 3\right) }:=\left\{ \xi ,\tau <\mathcal{O}\left(
\epsilon \right) \right\} .
\]%
Note that in order to reduce the original singular perturbed problem to a
set of simpler regular problems it is necessary to use inner (rescaled)
variable in the relevant subdomains. However, it is expedient to perform
this procedure during investigation of the boundary value problem (\ref{nd4}%
)-(\ref{nd9}) in appropriate subdomains of the diffusion boundary layer.

\subsection{General form of asymptotic solution}

Our aim is to find the leading-term asymptotic solution of the problem (\ref%
{nd4})-(\ref{nd9}) $u_{a}\left( \rho ,\xi ,\tau ;\epsilon \right) $
uniformly valid to order $\mathcal{O}\left( 1\right) $ in the whole domain $%
\Sigma _{\tau }$ (see Appendix).

It is convenient to divide the desired asymptotic solution $u$ in three
parts: outer $u^{\left( 0\right) }$ (regular in $\Omega ^{\left( 0\right) }$%
), boundary layer $u^{\left( b\right) }$ and corner boundary layer $%
u^{\left( c\right) }$. Thus the asymptotic solution $u$ may be sought in the
form 
\begin{equation}
u=u^{\left( 0\right) }+u^{\left( b\right) }+u^{\left( c\right) }.
\label{sd6}
\end{equation}%
In its turn the boundary layer solution is 
\[
u^{\left( b\right) }=u^{\left( 1\right) }+\widetilde{u}^{\left( 1\right)
}+u^{\left( 3\right) }, 
\]%
where $u^{\left( 1\right) }$ and $\widetilde{u}^{\left( 1\right) }$ are the
left and right boundary layer solutions given in $\Omega _{0}^{\left(
1\right) }$ and $\Omega _{1}^{\left( 1\right) }$, respectively; $u^{\left(
3\right) }$ is the initial boundary layer solution in $\Omega _{0}^{\left(
3\right) }$. Finally the corner boundary layer solution $u^{\left( c\right)
} $ naturally is divided into the sum 
\[
u^{\left( c\right) }=u^{\left( 2\right) }+\widetilde{u}^{\left( 2\right) }, 
\]%
where $u^{\left( 2\right) }$ and $\widetilde{u}^{\left( 2\right) }$ are the
left and right corner boundary layer solutions in $\Omega _{0}^{\left(
2\right) }$ and $\Omega _{1}^{\left( 2\right) }$, respectively. To find
explicit form of the above asymptotic solutions one should rewrite original
boundary value problem in corresponding subdomains using so-called stretched
(or inner) variables inherent in these subdomains (see Sec. V).

The relevant boundary and initial conditions must also take into account the
discrepancies arising for the boundary layer solutions $u^{\left( 1\right) }$%
, $\widetilde{u}^{\left( 1\right) }$ and $u^{\left( 3\right) }$ due to the
function $u^{\left( 0\right) }$. Furthermore the corner functions $u^{\left(
2\right) }$ and $\widetilde{u}^{\left( 2\right) }$ should correct
discrepancies caused by functions $u^{\left( 1\right) }$, $\widetilde{u}%
^{\left( 1\right) }$ and $u^{\left( 3\right) }$. This procedure is
represented by the diagram 
\begin{equation}
\begin{tabular}{lllll}
$u^{\left( 1\right) }$ & $\leftarrow $ & $u^{\left( 0\right) }$ & $%
\rightarrow $ & $\widetilde{u}^{\left( 1\right) }$ \\ 
$\downarrow $ &  & $\downarrow $ &  & $\downarrow $ \\ 
$u^{\left( 2\right) }$ & $\leftarrow $ & $u^{\left( 3\right) }$ & $%
\rightarrow $ & $\widetilde{u}^{\left( 2\right) }$%
\end{tabular}%
\ .  \label{sd7}
\end{equation}%
It is worth noting here that the above procedure is similar to that used in
the reflections method.\cite{TraytakCM:2007} On the other hand the matching
conditions describe exponentially small influence of the appropriate
solutions in the opposite directions 
\begin{equation}
\begin{tabular}{lllll}
$u^{\left( 1\right) }$ & $\rightarrow $ & $u^{\left( 0\right) }$ & $%
\leftarrow $ & $\widetilde{u}^{\left( 1\right) }$ \\ 
$\uparrow $ &  & $\uparrow $ &  & $\uparrow $ \\ 
$u^{\left( 2\right) }$ & $\rightarrow $ & $u^{\left( 3\right) }$ & $%
\leftarrow $ & $\widetilde{u}^{\left( 2\right) }$%
\end{tabular}%
\ .  \label{sd8}
\end{equation}%
The explicit form of functions included in (\ref{sd6}) will be found during
the asymptotic solution iterative procedure.

\section{Solution in the outer subdomain $\Omega ^{\left( 0\right) }$.}

\subsection{Zeroth-order outer approximation. The Fick-Jacobs equation}

According to common matched asymptotic expansions method technique consider
first the diffusion equation for slow variables $\xi $ and $\tau $ in the
outer subdomain $\Omega ^{\left( 0\right) }$.\cite{Lagerstrom:1988,Ilin:1992}
Let us look for the asymptotic solution to Eq. (\ref{nd4}) with conditions (%
\ref{nd6})-(\ref{nd9}) in $\Omega ^{\left( 0\right) }$ as a regular
perturbation expansion in the relaxation parameter 
\begin{equation}
u^{\left( 0\right) }\left( \rho ,\xi ,\tau \right)
=\sum\limits_{n=0}^{\infty }u_{n}^{\left( 0\right) }\left( \rho ,\xi ,\tau
\right) \epsilon ^{n}\text{ }\quad \text{as }\epsilon \rightarrow 0.
\label{t11}
\end{equation}%
So outer subdomain $\Omega ^{\left( 0\right) }$ sometimes is called regular
one. Note that, although here we limit ourselves by determination of the
leading order term $\mathcal{O}\left( 1\right) $, using the proposed
approach, one can find functions $u_{n}^{\left( 0\right) }\left( \rho ,\xi
,\tau \right) $ of any reasonable number $n$. Substitution of (\ref{t11}) in
Eq. (\ref{nd4}) leads to the following iterative equations 
\begin{equation}
\mathcal{L}_{\rho }u_{0}^{\left( 0\right) }=0,  \label{t12}
\end{equation}%
\begin{equation}
\mathcal{L}_{\rho }u_{n}^{\left( 0\right) }=-\mathcal{L}_{F}u_{n-1}^{\left(
0\right) },\quad n\geq 1  \label{t13}
\end{equation}%
and in its turn conditions (\ref{nd9}) read 
\begin{equation}
\left. u_{n}^{\left( 0\right) }\right\vert _{\rho =0}<\infty ,\qquad \left. 
\frac{\partial u_{n}^{\left( 0\right) }}{\partial \rho }\right\vert _{\rho
=0}=0,\quad n\geq 0.  \label{t13a}
\end{equation}%
Similarly, inserting (\ref{t11}) into the reflecting wall condition (\ref%
{nd8}), we get the following recurrence relations 
\begin{equation}
\left. \frac{\partial u_{0}^{\left( 0\right) }}{\partial \rho }\right\vert
_{\rho =R\left( \xi \right) }=0,  \label{t13b}
\end{equation}%
\begin{equation}
\left. \left[ \frac{\partial u_{n}^{\left( 0\right) }}{\partial \rho }-\frac{%
\partial u_{n-1}^{\left( 0\right) }}{\partial \xi }R^{^{\prime }}\left( \xi
\right) \right] \right\vert _{\rho =R\left( \xi \right) }=0,\quad n\geq 1.
\label{t13c}
\end{equation}

One can see that the general solution to the quasi steady-state Eq. (\ref%
{t12}) is 
\begin{equation}
u_{0}^{\left( 0\right) }\left( \rho ,\xi ,\tau \right) =u_{00}^{\left(
0\right) }\left( \xi ,\tau \right) +u_{01}^{\left( 0\right) }\left( \xi
,\tau \right) \ln \rho .  \label{t14}
\end{equation}%
Here $u_{00}^{\left( 0\right) }\left( \xi ,\tau \right) $ and $%
u_{01}^{\left( 0\right) }\left( \xi ,\tau \right) $ are unknown functions to
be determined from the boundary conditions (\ref{t13a}) and (\ref{t13c}).
With the aid of conditions (\ref{t13a}) we see that $u_{01}^{\left( 0\right)
}\left( \xi ,\tau \right) \equiv 0$ and therefore $u_{0}^{\left( 0\right)
}\left( \rho ,\xi ,\tau \right) =u_{00}^{\left( 0\right) }\left( \xi ,\tau
\right) $, which automatically obeys condition (\ref{t13b}).

Consider now the general iterative problem (\ref{t13}), (\ref{t13a}) and (%
\ref{t13c}) for $n\geq 1$. It is clear that this problem may be insoluble
because the unperturbed operator $\mathcal{L}_{\rho }$ is in spectrum \cite%
{Vishik:1960} (see Appendix). This circumstance leads to the fact that the
approximations in (\ref{t11}) cannot be given explicitly, and they are
determined by some unknown functions $u_{n}^{\left( 0\right) }\left( \rho
,\xi ,\tau \right) $. Let us find the solvability condition for the
iterative problem (\ref{t13}), (\ref{t13a}) and (\ref{t13c}). Multiplying
Eq. (\ref{t13}) by $\rho $ and integrating then with respect to $\rho $ from 
$\rho =0$ we arrive at 
\[
\int\limits_{0}^{\rho }\rho \mathcal{L}_{F}u_{n-1}^{\left( 0\right) }d\rho
=\rho \frac{\partial u_{n}^{\left( 0\right) }}{\partial \rho }.
\]%
Hence, utilizing the recurrence boundary condition (\ref{t13c}), the desired
solvability condition for $u_{n}^{\left( 0\right) }\left( \rho ,\xi ,\tau
\right) $ reads 
\begin{equation}
\int\limits_{0}^{R\left( \xi \right) }\rho \mathcal{L}_{F}u_{n-1}^{\left(
0\right) }d\rho =R\left( \xi \right) R^{\prime }\left( \xi \right) \left. 
\frac{\partial u_{n-1}^{\left( 0\right) }}{\partial \xi }\right\vert _{\rho
=R\left( \xi \right) },\quad n\geq 1.  \label{t15}
\end{equation}%
It is important to underline that solvability condition (\ref{t15})
eventually follows from the reflecting boundary condition (\ref{nd8})
imposed on the tube wall $\partial \Sigma _{w}$. In specific case at $n=1$
from (\ref{t15}) we get straightforwardly the following condition 
\begin{equation}
\mathcal{L}_{F}u_{0}^{\left( 0\right) }=2\frac{R^{\prime }\left( \xi \right) 
}{R\left( \xi \right) }\frac{\partial u_{0}^{\left( 0\right) }}{\partial \xi 
}.  \label{t17a}
\end{equation}%
One can readily see that obtained condition (\ref{t17a}) is a dimensionless
form of the FJE (see Sec. X).

For further treatment it is convenient to put down the dimensionless FJE (%
\ref{t17a}) in a compact form 
\begin{equation}
\mathcal{L}_{FJ}u_{0}^{\left( 0\right) }=0,  \label{t19a}
\end{equation}%
introducing the dimensionless Fick-Jacobs operator 
\[
\mathcal{L}_{FJ}:=\mathcal{L}_{F}-2\frac{R^{\prime }\left( \xi \right) }{%
R\left( \xi \right) }\frac{\partial }{\partial \xi }.
\]%
Moreover the zeroth-order approximation in the outer solution $u_{0}^{\left(
0\right) }\left( \xi ,\tau \right) $ we can naturally call the Fick-Jacobs
approximation (FJA). To complete the derivation of the FJA one needs to
infer the appropriate initial and boundary conditions using the asymptotic
solutions of the posed problem in the diffusion boundary layer $\Omega
^{\left( b\right) }$. Thus it follows from the above treatment that
mathematically the FJE is nothing other than a condition of solvability for
the function $u_{1}^{\left( 0\right) }\left( \rho ,\xi ,\tau \right) $. This
feature of $u_{1}^{\left( 0\right) }\left( \rho ,\xi ,\tau \right) $ is
common with the Hilbert approximation in the kinetic theory for low Knudsen
numbers (see Sec. VIII).

Assuming that (\ref{t19a}) holds true, it is clear that the general solution
to inhomogenious Eq. (\ref{t13}), which satisfies conditions (\ref{t13a}) is 
\begin{equation}
u_{1}^{\left( 0\right) }\left( \rho ,\xi ,\tau \right) =u_{10}^{\left(
0\right) }\left( \xi ,\tau \right) -\frac{1}{4}\rho ^{2}\mathcal{L}%
_{F}u_{0}^{\left( 0\right) },  \label{t16}
\end{equation}%
where $u_{10}^{\left( 0\right) }\left( \xi ,\tau \right) $ is an arbitrary
function to be found during asymptotic solution. Hence it is important to
note that the outer approximation of order $\mathcal{O}\left( \epsilon
\right) $ contains a term depending on the transversal variable $\rho $. 

To end this subsection, we observe that solution $u\left( \rho .\xi .\tau
\right) $ is nonanalytic in the relaxation parameter $\epsilon $ and,
therefore, regular expansion (\ref{t11}) in powers of $\epsilon $ fails to
give uniformly valid approximation in the whole domain $\Sigma _{\tau }$. In
this connection we note that expansion (\ref{t11}) is an analog of the
Hilbert expansion for solution of the Boltzmann equation (see Sec. VIII for
details). Thus, as we mentioned in Sec. III, to find the uniformly valid
approximation, one has to solve appropriate boundary value problems
concerning the diffusion boundary layers.

\subsection{Forms of the Fick-Jacobs equation}

Analysis of the literature showed that the classical form of FJE (\ref{in1})
(or in case of anisotropic diffusion (\ref{sp7})) is the most common in
theoretical and applied papers. However we believe that the most natural
form of the FJE for the axially symmetric tubes is the following divergent
form \cite{Kalnay2:2005} 
\begin{equation}
\frac{\partial u_{0}^{\left( 0\right) }}{\partial \tau }=\frac{1}{R\left(
\xi \right) ^{2}}\frac{\partial }{\partial \xi }\left[ R\left( \xi \right)
^{2}\frac{\partial u_{0}^{\left( 0\right) }}{\partial \xi }\right] .
\label{di4}
\end{equation}%
It seems interesting that the right hand side of Eq. (\ref{di4}) resembles
the Laplacian action in conformally flat metric, which was rather widely
investigated in theoretical physics.\cite{Turbiner:1984} The connection of
this equation with classical FJE (\ref{sp7}) is known.\cite{Kalnay2:2005}\
Utilizing the cylindrical coordinates we can write relation (\ref{in2}) as
follows: 
\begin{equation}
c\left( z,t\right) =2\pi \int\limits_{0}^{r\left( z\right) }C\left(
r,z,t\right) rdr  \label{di5}
\end{equation}%
and making notation 
\begin{equation}
C_{M}u_{0}^{\left( 0\right) }\left( z,t\right) =c_{0}^{\left( 0\right)
}\left( z,t\right) :=\frac{1}{A\left( z\right) }\lim_{\epsilon \rightarrow
0}c\left( z,t\right)  \label{di6}
\end{equation}%
we obtain the FJE in the classical form (\ref{in1}).

The FJE in the form (\ref{di4}) may be useful, g.e., to describe diffusion
in a long conical tube of the radius given by linear function%
\begin{equation}
R\left( \xi \right) =a\xi +b,  \label{di7}
\end{equation}%
where $a$ and $b$ are some constants. Indeed, assuming for definiteness that 
$a$ and $b$ are positive, by means of substitution (\ref{di7}) we can reduce
the FJE (\ref{di4}) to well-known spherically symmetric diffusion equation 
\begin{equation}
\frac{\partial u_{0}^{\left( 0\right) }}{\partial \tau }=a^{2}\frac{1}{R^{2}}%
\frac{\partial }{\partial R}\left[ R^{2}\frac{\partial u_{0}^{\left(
0\right) }}{\partial R}\right]  \label{di8}
\end{equation}%
in a hollow sphere $b<R<a+b$.\cite{Carslaw:1959}

To present one more example rewrite Eq. (\ref{di4}) in the dimensional form 
\begin{equation}
R^{2}\left( z\right) \frac{\partial c_{0}^{\left( 0\right) }}{\partial t}%
=D_{\Vert }\frac{\partial }{\partial z}\left[ R^{2}\left( z\right) \frac{%
\partial c_{0}^{\left( 0\right) }}{\partial z}\right] .  \label{dis3}
\end{equation}%
One can see that multiplication of (\ref{dis3}) by $\pi $ and integration
with respect to $z$ from $0$ to any current $z$ gives 
\begin{equation}
\Phi \left( z,t\right) =\Phi \left( 0,t\right) -\int\limits_{0}^{z}\frac{%
\partial c_{0}^{\left( 0\right) }}{\partial t}dV,  \label{dis4}
\end{equation}%
where $dV=\pi R^{2}\left( z\right) dz$ and 
\[
\Phi \left( z,t\right) :=-D_{\Vert }\pi R^{2}\left( z\right) \frac{\partial 
}{\partial z}c_{0}^{\left( 0\right) }\left( z,t\right) 
\]%
is the total flux of diffusing particles through the cross section at point $%
z$. For the steady state flux $\Phi _{s}\left( z\right) $ ($t\gg t_{L}$)
relationship (\ref{dis4}) takes the simplest form 
\begin{equation}
\Phi _{s}\left( z\right) =\Phi _{s}\left( 0\right) .  \label{dis5}
\end{equation}%
Thus, the FJE is similar to continuity equation and Eq. (\ref{dis5}) is an
analog to known Bernoulli's principle in ideal fluid dynamics. The latter
fact supports the analogy between the FJA $u_{0}^{\left( 0\right) }$ and the
Hilbert solution to the Boltzmann equation at low Knudsen numbers (see Sec.
IV and below).

Eq. (\ref{t19a}) represents one more convenient form of the FJE, which for
dimensional variables reads 
\begin{equation}
\frac{\partial c_{0}^{\left( 0\right) }}{\partial \tau }-D_{\parallel }\frac{%
\partial ^{2}c_{0}^{\left( 0\right) }}{\partial z^{2}}=V\left( z\right) 
\frac{\partial c_{0}^{\left( 0\right) }}{\partial z}.  \label{dis6}
\end{equation}%
Here we denote $V\left( z\right) :=V_{L}A^{-1}dA/dz$ the effective drift
velocity of diffusing particles along the $z$ axis, and $V_{L}=D_{\parallel
}/L$ the characteristic longitudinal diffusion velocity. It seems that
convective diffusion interpretation (\ref{dis6}) appeared to be even more
appropriate for investigation of the FJE than widely used entropy potential
form (\ref{in2a}). This ensues from the fact that nowadays mathematical
theory of convective diffusion equation is thoroughly elaborated in all
aspects.

\section{Solution in the subdomains $\Omega _{0}^{\left( 1\right) }$ and $%
\Omega _{1}^{\left( 1\right) }$. Boundary conditions for the Fick-Jacobs
equation}

Let us study the solution of the problem (\ref{nd4})-(\ref{nd9}) in the
spatial diffusion boundary layer subdomains $\Omega _{0}^{\left( 1\right) }$
and $\Omega _{1}^{\left( 1\right) }$ attached to the endpoints ($\xi
=\left\{ 0,1\right\} $) (see Fig.~\ref{fig:dom}). In subdomains $\Omega _{0}^{\left(
1\right) }$ and $\Omega _{1}^{\left( 1\right) }$ we introduce so-called
inner coordinates: new stretched spatial variables $\xi ^{\ast }=\xi /\sqrt{%
\epsilon }$ and $\widetilde{\xi }=\left( 1-\xi \right) /\sqrt{\epsilon }$,
respectively, leaving slow time $\tau $ unscaled. So in the left and right
subdomains $\Omega _{0}^{\left( 1\right) }$ and $\Omega _{1}^{\left(
1\right) }$ one has $\xi ^{\ast }=\mathcal{O}\left( 1\right) $ and $%
\widetilde{\xi }=\mathcal{O}\left( 1\right) $ as $\epsilon \rightarrow 0$.
Asymptotic solutions to the problem (\ref{nd4})-(\ref{nd9}) behave similarly
in $\Omega _{0}^{\left( 1\right) }$ and $\Omega _{1}^{\left( 1\right) }$,
therefore, for definiteness we consider in detail the solution corresponding
to the left subdomain $\Omega _{0}^{\left( 1\right) }$ only. Rewriting the
boundary value problem (\ref{nd4})-(\ref{nd9}) in the inner coordinates $%
\left( \rho ,\xi ^{\ast },\tau \right) $ of $\Omega _{0}^{\left( 1\right) }$
for the inner solution $u^{\left( 1\right) }\left( \rho ,\xi ^{\ast };\tau
\right) $ we obtain 
\begin{equation}
\frac{\partial ^{2}u^{\left( 1\right) }}{\partial \xi ^{\ast 2}}-\mathcal{L}%
_{\rho }u^{\left( 1\right) }=\epsilon \frac{\partial u^{\left( 1\right) }}{%
\partial \tau }\qquad \text{ in }\Omega _{0}^{\left( 1\right) },  \label{bc1}
\end{equation}%
\begin{equation}
\left. u^{\left( 1\right) }\right\vert _{\rho =0}<\infty ,\qquad \left. 
\frac{\partial u^{\left( 1\right) }}{\partial \rho }\right\vert _{\rho =0}=0,
\label{bc3}
\end{equation}%
\begin{equation}
\left. \left[ \frac{\partial u^{\left( 1\right) }\left( \rho ,\xi ^{\ast
};\tau \right) }{\partial \rho }-\sqrt{\epsilon }\frac{\partial u^{\left(
1\right) }}{\partial \xi ^{\ast }}R^{^{\prime }}\left( \sqrt{\epsilon }\xi
^{\ast }\right) \right] \right\vert _{\rho =R\left( \sqrt{\epsilon }\xi
^{\ast }\right) }=0.  \label{bc4}
\end{equation}

Let us observe that conditions on the $z$ axis (\ref{bc3}) must be satisfied
in all other subdomains of the boundary layer ($\Omega _{1}^{\left( 1\right)
}$, $\Omega _{0}^{\left( 2\right) }$, $\Omega _{1}^{\left( 2\right) }$, and $%
\Omega _{0}^{\left( 3\right) }$), so for brief henceforward we omit them
later in the text. Due to the type of Eq. (\ref{bc1}) subdomain $\Omega
_{0}^{\left( 1\right) }$ ($\Omega _{1}^{\left( 1\right) }$) is often termed
as elliptic boundary layer.\cite{Ilin:1992}

Find now the appropriate boundary conditions for $u^{\left( 1\right) }$.
Bearing in mind the derivation of uniformly valid approximation (\ref{sd6}),
consider the partial sum 
\begin{equation}
u^{\left( 0,1\right) }\left( \rho ,\xi ,\xi ^{\ast },\tau \right) =u^{\left(
0\right) }\left( \rho ,\xi ,\tau \right) +u^{\left( 1\right) }\left( \rho
,\xi ^{\ast };\tau \right)   \label{bc4a}
\end{equation}%
which is defined in $\Omega ^{\left( 0\right) }\cup \Omega _{0}^{\left(
1\right) }$ with appropriate matching conditions. It is worth noting that
both outer $u^{\left( 0\right) }\left( \rho ,\xi ,\tau \right) $ and inner $%
u^{\left( 1\right) }\left( \rho ,\xi ^{\ast };\tau \right) $ solutions are
approximations to the same solution but defined in outer $\Omega ^{\left(
0\right) }$ and inner $\Omega _{0}^{\left( 1\right) }$ subdomains,
respectively. So the compound approximation $u^{\left( 0,1\right) }$ should
obeys the left boundary condition (\ref{nd7}) and satisfies the
corresponding diffusion equation (\ref{nd4}) in $\Omega ^{\left( 0\right)
}\cup \Omega _{0}^{\left( 1\right) }$. Substitution of approximation (\ref%
{bc4a}) into Eq. (\ref{nd4}) yields 
\[
\left( \mathcal{L}_{\rho }+\epsilon \mathcal{L}_{F}\right) u^{\left(
0\right) }\left( \rho ,\xi ,\tau \right) 
\]%
\[
=\left( \frac{\partial ^{2}}{\partial \xi ^{\ast 2}}-\mathcal{L}_{\rho
}-\epsilon \frac{\partial }{\partial \tau }\right) u^{\left( 1\right)
}\left( \rho ,\xi ^{\ast };\tau \right) .
\]%
Hence, employing the fact, that $\xi $ and $\xi ^{\ast }$ are independent
variables, we obtain Eqs (\ref{nd4}) and (\ref{bc1}). In its turn from the
left boundary condition (\ref{nd7}) we have 
\begin{equation}
\left. u^{\left( 1\right) }\right\vert _{\xi ^{\ast }=0}=g_{1}\left( \rho
,\tau \right) -\left. u^{\left( 0\right) }\left( \rho ,\xi ,\tau \right)
\right\vert _{\xi =0}.  \label{bc4b}
\end{equation}%
Missing right boundary condition for $u^{\left( 1\right) }$ may be found
using the matching condition between inner and outer solutions \cite%
{Ilin:1992}, i.e. 
\[
\left. u^{\left( 0,1\right) }\left( \rho ,\xi ,\xi ^{\ast },\tau \right)
\right\vert _{\xi ^{\ast }\rightarrow \infty }\rightarrow u^{\left( 0\right)
}\left( \rho ,\xi ,\tau \right) 
\]%
that immediately leads to the desired boundary condition 
\begin{equation}
\left. u^{\left( 1\right) }\left( \rho ,\xi ^{\ast };\tau \right)
\right\vert _{\xi ^{\ast }\rightarrow \infty }\rightarrow 0.  \label{bc5}
\end{equation}%
Note that hereafter the limit as $\xi ^{\ast }\rightarrow \infty $ means
that $\epsilon \rightarrow 0$ provided $\xi $ is fixed.

It is clear that inside $\Omega _{0}^{\left( 1\right) }$ the problem under
study becomes regular, so we can seek the solution $u^{\left( 1\right)
}\left( \rho ,\xi ^{\ast },\tau \right) $ in the form of the expansion 
\begin{equation}
u^{\left( 1\right) }\left( \rho ,\xi ^{\ast },\tau \right)
=\sum\limits_{m=0}^{\infty }u_{m}^{\left( 1\right) }\left( \rho ,\xi ^{\ast
};\tau \right) \epsilon ^{m/2}\text{ }\quad \text{as }\epsilon \rightarrow 0,
\label{bc6}
\end{equation}%
where $u_{m}^{\left( 1\right) }\left( \rho ,\xi ^{\ast };\tau \right) $ are
so-called functions of the boundary layer.\cite{Ilin:1992} One can see that
implicitly the dependence on slow time $\tau $ arises in the second order
approximation ($m=2$) only. In this way we formally reduced the problem in
the inner subdomain $\Omega _{0}^{\left( 1\right) }$ to the quasi
steady-state problem (we have only parametric dependence upon time $\tau $)
posed on the semi-infinite tube bounded at $\xi ^{\ast }=0$.

Hence for the zeroth-order approximation we gain the problem 
\begin{equation}
\frac{\partial ^{2}u_{0}^{\left( 1\right) }}{\partial \xi ^{\ast 2}}-%
\mathcal{L}_{\rho }u_{0}^{\left( 1\right) }=0,\qquad 0<\rho <R_{0},
\label{bc7}
\end{equation}%
\begin{equation}
\left. u_{0}^{\left( 1\right) }\right\vert _{\xi ^{\ast }=0}=g_{1}\left(
\rho ,\tau \right) -\left. u_{0}^{\left( 0\right) }\right\vert _{\xi =0},
\label{bc8}
\end{equation}%
\begin{equation}
\left. u_{0}^{\left( 1\right) }\right\vert _{\xi ^{\ast }\rightarrow \infty
}\rightarrow 0,  \label{bc8a}
\end{equation}%
\begin{equation}
\left. \frac{\partial u_{0}^{\left( 1\right) }}{\partial \rho }\right\vert
_{\rho =R_{0}}=0.  \label{bc9}
\end{equation}%
Therefore for $u_{0}^{\left( 1\right) }$ the reflecting wall condition (\ref%
{bc4}) simplifies to the relevant condition (\ref{bc9}) on the circular
cylinder of constant radius $R_{0}:=R\left( 0\right) $.

The general solution to Eq. (\ref{bc7}) satisfying the reflecting condition (%
\ref{bc9}) is 
\begin{equation}
u_{0}^{\left( 1\right) }\left( \rho ,\xi ^{\ast };\tau \right)
=\sum\limits_{k=0}^{\infty }b_{k}\left( \tau \right) e^{-\lambda _{k}\xi
^{\ast }}\widehat{J}_{0}\left( \lambda _{k}\frac{\rho }{R_{0}}\right) ,
\label{bc10}
\end{equation}%
where $\left\{ \widehat{J}_{0}\left( \lambda _{k}\rho /R_{0}\right) \right\}
_{k=0}^{\infty }$ is the complete orthonormal system defined in Appendix.

Matching condition (\ref{bc8a}) leads to $b_{0}\left( \tau \right) =0$ that
yields the desired boundary condition for the solution of the FJE $%
u_{0}^{\left( 0\right) }\left( \xi ,\tau \right) $ at the left endpoint ($%
\xi =0$) 
\begin{equation}
\left. u_{0}^{\left( 0\right) }\left( \xi ,\tau \right) \right\vert _{\xi
=0}=\left\langle g_{1},\widehat{J}_{0}\left( 0\right) \right\rangle _{%
\mathcal{H}_{0}}\widehat{J}_{0}\left( 0\right) =\frac{2}{R_{0}^{2}}%
\int\limits_{0}^{R_{0}}\rho g_{1}\left( \rho ,\tau \right) d\rho .
\label{bc11}
\end{equation}%
For $k\geq 1$ unknown coefficients $b_{k}\left( \tau \right) $ are 
\begin{equation}
b_{k}\left( \tau \right) =\left\langle g_{1},\widehat{J}_{0}\left( \lambda
_{k}\frac{\rho }{R_{0}}\right) \right\rangle _{\mathcal{H}_{0}}.
\label{bc12}
\end{equation}%
Similar treatment of the inner solution in the right subdomain $\Omega
_{1}^{\left( 1\right) }$ gives 
\begin{equation}
\widetilde{u}_{0}^{\left( 1\right) }\left( \rho ,\widetilde{\xi };\tau
\right) =\sum\limits_{k=0}^{\infty }\widetilde{b}_{k}\left( \tau \right)
e_{k}^{-\lambda _{k}\widetilde{\xi }}\widehat{J}_{0}\left( \lambda _{k}\frac{%
\rho }{R_{1}}\right) ,  \label{bc13}
\end{equation}%
Hence $\widetilde{b}_{0}\left( \tau \right) =0$ and the second boundary
condition for the FJA $u_{0}^{\left( 0\right) }\left( \xi ,\tau \right) $ at
the right endpoint ($\xi =1$) is 
\begin{equation}
\left. u_{0}^{\left( 0\right) }\left( \xi ,\tau \right) \right\vert _{\xi
=1}=\left\langle g_{2},\widehat{J}_{0}\left( 0\right) \right\rangle _{%
\mathcal{H}_{1}}\widehat{J}_{0}\left( 0\right) =\frac{2}{R_{1}^{2}}%
\int\limits_{0}^{R_{1}}\rho g_{2}\left( \rho ,\tau \right) d\rho ,
\label{bc14}
\end{equation}%
where $R_{1}:=R\left( 1\right) $ and 
\begin{equation}
\widetilde{b}_{k}\left( \tau \right) =\left\langle g_{2},\widehat{J}%
_{0}\left( \lambda _{k}\frac{\rho }{R_{1}}\right) \right\rangle _{\mathcal{H}%
_{1}},\quad k\geq 1.  \label{bc15}
\end{equation}

\section{Solution in the subdomain $\Omega _{0}^{\left( 3\right) }$.
Initial conditions for the Fick-Jacobs equation}

Now we dwell on the solution to problem (\ref{nd4})-(\ref{nd9}) in the
initial diffusion boundary layer subdomain $\Omega _{0}^{\left( 3\right) }$
attached to the the initial time $\tau =0$ (see Fig.~\ref{fig:dom}). In this subdomain
inner variables are $\xi $ and the stretched (fast) time $\tau ^{\ast }=\tau
/\epsilon $ ($\tau ^{\ast }=\mathcal{O}\left( 1\right) $ as $\epsilon
\rightarrow 0$). Using these variables in the original problem (\ref{nd4})-(%
\ref{nd9}) similarly to the previous case we arrive at 
\begin{equation}
\frac{\partial u^{\left( 3\right) }}{\partial \tau ^{\ast }}+\mathcal{L}%
_{\rho }u^{\left( 3\right) }=\epsilon \frac{\partial ^{2}u^{\left( 3\right) }%
}{\partial \xi ^{2}},  \label{ic1}
\end{equation}%
\begin{equation}
\left. u^{\left( 3\right) }\right\vert _{\tau ^{\ast }=0}=g_{0}\left( \rho
,\xi \right) -\left. u^{\left( 0\right) }\left( \rho ,\xi ,\tau \right)
\right\vert _{\tau =0},  \label{ic2}
\end{equation}%
\begin{equation}
\left. \left[ \frac{\partial u^{\left( 3\right) }}{\partial \rho }-\epsilon 
\frac{\partial u^{\left( 3\right) }}{\partial \xi }R^{^{\prime }}\left( \xi
\right) \right] \right\vert _{\rho =R\left( \xi \right) }=0.  \label{ic3}
\end{equation}%
One can see that now we obtained effectively the boundary value problem for
the infinitely long solid of revolution. According to the type of Eq. (\ref%
{ic1}) the subdomain $\Omega _{0}^{\left( 3\right) }$ is often called as
parabolic boundary layer.\cite{Ilin:1992} We also must add to (\ref{ic2})
and (\ref{ic3}) the matching condition for the inner solution 
\begin{equation}
\left. u^{\left( 3\right) }\left( \rho ,\tau ^{\ast };\xi \right)
\right\vert _{\tau ^{\ast }\rightarrow \infty }\rightarrow 0.  \label{ic4}
\end{equation}

The corresponding regular expansion of the inner solution in $\Omega
_{0}^{\left( 3\right) }$ reads

\begin{equation}
u^{\left( 3\right) }\left( \rho ,\xi ,\tau ^{\ast }\right)
=\sum\limits_{m=0}^{\infty }u_{m}^{\left( 3\right) }\left( \rho ,\tau ^{\ast
};\xi \right) \epsilon ^{m}\text{ }\quad \text{as }\epsilon \rightarrow 0.
\label{ic5}
\end{equation}%
Here the functions of the boundary layer $u_{m}^{\left( 3\right) }\left(
\rho ,\tau ^{\ast };\xi \right) $ depend upon $\xi $ as a parameter. So for
the zeroth-order function $u_{0}^{\left( 3\right) }\left( \rho ,\tau ^{\ast
};\xi \right) $ we get the following boundary value problem 
\begin{equation}
\frac{\partial u_{0}^{\left( 3\right) }}{\partial \tau ^{\ast }}+\mathcal{L}%
_{\rho }u_{0}^{\left( 3\right) }=0,  \label{ic6}
\end{equation}%
\begin{equation}
\left. u_{0}^{\left( 3\right) }\right\vert _{\tau ^{\ast }=0}=g_{0}\left(
\rho ,\xi \right) -\left. u_{0}^{\left( 0\right) }\right\vert _{\tau =0},
\label{ic7}
\end{equation}%
\begin{equation}
\left. \frac{\partial u_{0}^{\left( 3\right) }}{\partial \rho }\right\vert
_{\rho =R\left( \xi \right) }=0.  \label{ic8}
\end{equation}%
One can see that the obtained problem (\ref{ic6})-(\ref{ic8}) describes the
diffusion into the infinite circular cylinder of radius $R\left( \xi \right) 
$. It is clear that general solution to Eq. (\ref{ic6}), which obey the
reflecting condition (\ref{ic8}), may be expressed as 
\begin{equation}
u_{0}^{\left( 3\right) }\left( \rho ,\tau ^{\ast };\xi \right)
=\sum\limits_{k=0}^{\infty }a_{k}\left( \xi \right) e^{-\lambda _{k}^{2}\tau
^{\ast }}\widehat{J}_{0}\left( \lambda _{k}\frac{\rho }{R\left( \xi \right) }%
\right) .  \label{ic9}
\end{equation}%
Setting $\tau ^{\ast }=0$, this implies that to satisfy the matching
condition 
\begin{equation}
\left. u_{0}^{\left( 3\right) }\right\vert _{\tau ^{\ast }\rightarrow \infty
}\rightarrow 0  \label{ic10}
\end{equation}%
we should impose $a_{0}\left( \xi \right) =0$ at that, utilizing initial
condition (\ref{ic7}), we have the desired initial condition for the FJA 
\begin{equation}
\left. u_{0}^{\left( 0\right) }\left( \xi ,\tau \right) \right\vert _{\tau
=0}=\left\langle g_{0},\widehat{J}_{0}\left( 0\right) \right\rangle _{%
\mathcal{H}_{\xi }}\widehat{J}_{0}\left( 0\right) =\frac{2}{R^{2}\left( \xi
\right) }\int\limits_{0}^{R\left( \xi \right) }\rho g_{0}\left( \rho ,\xi
\right) d\rho .  \label{ic11}
\end{equation}%
and expression for unknown coefficients $a_{k}\left( \xi \right) $ ($k\geq 1$%
)%
\begin{equation}
a_{k}\left( \xi \right) =\left\langle g_{0},\widehat{J}_{0}\left( \lambda
_{k}\frac{\rho }{R\left( \xi \right) }\right) \right\rangle _{\mathcal{H}%
_{\xi }}.  \label{ic12}
\end{equation}%
For the problem under study this formula gives the answer to the question
posed by Kalnay and Percus: "Having projected 2D equation to the 1D one ...,
one may ask the question: how then is the projected 1D initial density $%
P\left( x,0\right) $ related to the original $\rho \left( x,y,0\right) $,
and is there some reasonable projection algorithm?"\cite{Kalnay:2005}

\section{Solution in the corner subdomains $\Omega _0^{\left( 2\right)
} $ and $\Omega _1^{\left( 2\right) }$}

It is clear that behavior of inner solution in the corner subdomains $\Omega
_0^{\left( 2\right) }$ and $\Omega _1^{\left( 2\right) }$ (see Fig.~\ref{fig:dom}) is
similar, therefore, for briefness we give the detailed treatment of the
relevant boundary value problem only in $\Omega _0^{\left( 2\right) }$. For
this purpose we define inner variables $(\xi ^{*},\tau ^{*})$ and rewrite
Eq. (\ref{nd4}) for the corner boundary layer function $u^{\left( 2\right)
}\left( \rho ,\xi ^{*},\tau ^{*}\right) $ as

\begin{equation}
\frac{\partial ^{2}u^{\left( 2\right) }}{\partial \xi ^{\ast 2}}-\mathcal{L}%
_{\rho }u^{\left( 2\right) }=\frac{\partial u^{\left( 2\right) }}{\partial
\tau ^{\ast }},\quad 0<\rho <R_{0}.  \label{mc1}
\end{equation}%
Similarly to the previous case the subdomain $\Omega _{0}^{\left( 2\right) }$
(or $\Omega _{1}^{\left( 2\right) }$) is also called as parabolic boundary
layer.\cite{Ilin:1992} The reflecting wall condition (\ref{nd8}) in $\Omega
_{0}^{\left( 2\right) }$ takes the form 
\begin{equation}
\left. \left[ \frac{\partial u^{\left( 2\right) }\left( \rho ,\xi ^{\ast
},\tau ^{\ast }\right) }{\partial \rho }-\sqrt{\epsilon }\frac{\partial
u^{\left( 2\right) }}{\partial \xi ^{\ast }}R^{^{\prime }}\left( \sqrt{%
\epsilon }\xi ^{\ast }\right) \right] \right\vert _{\rho =R\left( \sqrt{%
\epsilon }\xi ^{\ast }\right) }=0.  \label{mc2}
\end{equation}%
One can see immediately from geometry of the problem that desired solution $%
u^{\left( 2\right) }\left( \rho ,\xi ^{\ast },\tau ^{\ast }\right) $ does
not affect directly to the outer solution $u^{\left( 0\right) }\left( \xi
,\tau \right) $ in $\Omega ^{\left( 0\right) }$. According to scheme (\ref%
{sd7}) function $u^{\left( 2\right) }\left( \rho ,\xi ^{\ast },\tau ^{\ast
}\right) $ should be matched with $u^{\left( 1\right) }\left( \rho ,\xi
^{\ast },\tau \right) $ and $u^{\left( 3\right) }\left( \rho ,\xi ,\tau
^{\ast }\right) $ in order to correct discrepancies due to these functions
for initial and boundary conditions, respectively 
\begin{equation}
\left. u^{\left( 2\right) }\right\vert _{\tau ^{\ast }=0}=-\left. u^{\left(
1\right) }\right\vert _{\tau =0},\qquad \left. u^{\left( 2\right)
}\right\vert _{\xi ^{\ast }=0}=-\left. u^{\left( 3\right) }\right\vert _{\xi
=0}.  \label{mc3}
\end{equation}%
The relevant matching conditions (see scheme (\ref{sd8})) for the corner
boundary layer function $u^{\left( 2\right) }\left( \rho ,\xi ^{\ast },\tau
^{\ast }\right) $ in $\Omega _{0}^{\left( 1\right) }$ and $\Omega
_{0}^{\left( 3\right) }$ are 
\begin{equation}
\left. u^{\left( 2\right) }\right\vert _{\tau ^{\ast }\rightarrow \infty
}\rightarrow 0,\qquad \left. u^{\left( 2\right) }\right\vert _{\xi ^{\ast
}\rightarrow \infty }\rightarrow 0.  \label{mc5}
\end{equation}%
One can see that obtained problem (\ref{mc1})-(\ref{mc5}) effectively
describes the time-dependent diffusion in a semi-infinite circular cylinder
of radius $R_{0}$.

Inside the corner boundary layer subdomain at issue $\Omega _{0}^{\left(
2\right) }$ we can seek the solution in the regular form 
\begin{equation}
u^{\left( 2\right) }\left( \rho ,\xi ^{\ast },\tau ^{\ast }\right)
=\sum\limits_{m=0}^{\infty }u_{m}^{\left( 2\right) }\left( \rho ,\xi ^{\ast
},\tau ^{\ast }\right) \epsilon ^{m/2}\text{ }\quad \text{as }\epsilon
\rightarrow 0.  \label{mc7}
\end{equation}%
Employing expressions (\ref{ic9}) and (\ref{bc10}) it is clear that the
zeroth-order approximation $u_{0}^{\left( 2\right) }$ to the corner function 
$u^{\left( 2\right) }$ in $\Omega _{0}^{\left( 2\right) }$ governs by the
problem 
\begin{equation}
\frac{\partial ^{2}u_{0}^{\left( 2\right) }}{\partial \xi ^{\ast 2}}-%
\mathcal{L}_{\rho }u_{0}^{\left( 2\right) }=\frac{\partial u_{0}^{\left(
2\right) }}{\partial \tau ^{\ast }},\quad 0<\rho <R_{0},  \label{mc8}
\end{equation}%
\begin{equation}
\left. u_{0}^{\left( 2\right) }\right\vert _{\tau ^{\ast
}=0}=-\sum\limits_{k=1}^{\infty }b_{k}\left( 0\right) e^{-\lambda _{k}\xi
^{\ast }}\widehat{J}_{0}\left( \lambda _{k}\frac{\rho }{R_{0}}\right) ,
\label{mc9}
\end{equation}%
\begin{equation}
\left. u_{0}^{\left( 2\right) }\right\vert _{\xi ^{\ast
}=0}=-\sum\limits_{k=1}^{\infty }a_{k}\left( 0\right) e^{-\lambda
_{k}^{2}\tau ^{\ast }}\widehat{J}_{0}\left( \lambda _{k}\frac{\rho }{R_{0}}%
\right) ,  \label{mc10}
\end{equation}%
\begin{equation}
\left. u_{0}^{\left( 2\right) }\right\vert _{\xi ^{\ast }\rightarrow \infty
}\rightarrow 0,  \label{mc11}
\end{equation}%
\begin{equation}
\left. \frac{\partial u_{0}^{\left( 2\right) }}{\partial \rho }\right\vert
_{\rho =R_{0}}=0.  \label{mc12}
\end{equation}%
It is convenient to look for solution of the obtained boudary value problem (%
\ref{mc8})-(\ref{mc12}) by means of projection method with respect to the
orthonormal basis $\left\{ \widehat{J}_{0}\left( \lambda _{k}\frac{\rho }{%
R_{0}}\right) \right\} _{k=0}^{\infty }$ (see Appendix). Multipling Eq. (\ref%
{mc8}) by $\widehat{J}_{0}\left( \lambda _{k}\frac{\rho }{R_{0}}\right) $
and using formulae (\ref{Hs2a}) and (\ref{Hs3})\ one gets%
\[
u_{0k}^{\left( 2\right) }\equiv 0,\qquad k=0;
\]%
\[
\frac{\partial ^{2}u_{0k}^{\left( 2\right) }}{\partial \xi ^{\ast 2}}%
-\lambda _{k}^{2}u_{0k}^{\left( 2\right) }=\frac{\partial u_{0k}^{\left(
2\right) }}{\partial \tau ^{\ast }},\qquad k\geq 1;
\]%
where 
\[
u_{0k}^{\left( 2\right) }\left( \xi ^{\ast },\tau ^{\ast }\right)
=\left\langle u_{0}^{\left( 2\right) },\widehat{J}_{0}\left( \lambda _{k}%
\frac{\rho }{R_{0}}\right) \right\rangle _{\mathcal{H}_{0}}.
\]%
Introducing for $k\geq 1$ a subsidiary function 
\[
w_{k}\left( \xi ^{\ast },\tau ^{\ast }\right) =e^{\lambda _{k}^{2}\tau
^{\ast }}u_{0k}^{\left( 2\right) }\left( \xi ^{\ast },\tau ^{\ast }\right) 
\]%
we finally reduce problem (\ref{mc8})-(\ref{mc12}) to%
\begin{equation}
\frac{\partial ^{2}w_{k}}{\partial \xi ^{\ast 2}}=\frac{\partial w_{k}}{%
\partial \tau ^{\ast }},  \label{mc14}
\end{equation}%
\begin{equation}
\left. w_{k}\right\vert _{\tau ^{\ast }=0}=-b_{k}\left( 0\right) e^{-\lambda
_{k}\xi ^{\ast }},  \label{mc14a}
\end{equation}%
\begin{equation}
\left. w_{k}\right\vert _{\xi ^{\ast }=0}=-a_{k}\left( 0\right) ,\quad
\left. w_{k}\right\vert _{\xi ^{\ast }\rightarrow \infty }\rightarrow 0.
\label{mc15}
\end{equation}%
One can easily derive that solution to the boundary value problem (\ref{mc14}%
)-(\ref{mc15}) reads \cite{Carslaw:1959}%
\[
w_{k}\left( \xi ^{\ast },\tau ^{\ast }\right) =-a_{k}\left( 0\right) %
\mbox{erfc}\left( \frac{\xi ^{\ast }}{2\sqrt{\tau ^{\ast }}}\right) 
\]%
\[
-\frac{1}{2}b_{k}\left( 0\right) e^{\lambda _{k}^{2}\tau ^{\ast }}\left[
e^{-\lambda _{k}\xi ^{\ast }}\mbox{erfc}\left( \lambda _{k}\sqrt{\tau ^{\ast
}}-\frac{\xi ^{\ast }}{2\sqrt{\tau ^{\ast }}}\right) \right. 
\]%
\begin{equation}
\left. -e^{\lambda _{k}\xi ^{\ast }}\mbox{erfc}\left( \lambda _{k}\sqrt{\tau
^{\ast }}+\frac{\xi ^{\ast }}{2\sqrt{\tau ^{\ast }}}\right) \right] ,
\label{mc16}
\end{equation}%
where 
\[
\mbox{erfc}\left( \varsigma \right) =\frac{2}{\sqrt{\pi }}\int_{\varsigma
}^{\infty }e^{-\alpha ^{2}}d\alpha 
\]%
is the complementary error function. Hence for the zeroth-order corner
function we have expansion%
\begin{equation}
u_{0}^{\left( 2\right) }\left( \rho ,\xi ^{\ast },\tau ^{\ast }\right)
=\sum\limits_{k=1}^{\infty }w_{k}\left( \xi ^{\ast },\tau ^{\ast }\right)
e^{-\lambda _{k}^{2}\tau ^{\ast }}\widehat{J}_{0}\left( \lambda _{k}\frac{%
\rho }{R_{0}}\right) .  \label{mc17}
\end{equation}%
It is obvious that function (\ref{mc17}) also satisfies matching condition (%
\ref{mc5}) as $\tau ^{\ast }\rightarrow \infty $.

The appropriate leading-term approximation for the inner solution $%
\widetilde{u}^{\left( 2\right) }\left( \rho ,\widetilde{\xi },\tau ^{\ast
}\right) $ in the corner subdomain $\Omega _{1}^{\left( 2\right) }$ may be
founded with the help of above derivation. For this purpose in (\ref{mc7})-(%
\ref{mc17}) one should implement the following substitutions%
\begin{equation}
\xi ^{\ast }\rightarrow \widetilde{\xi },\quad R_{0}\rightarrow R_{1},\quad
a_{k}\left( 0\right) \rightarrow a_{k}\left( 1\right) ,\quad b_{k}\left(
0\right) \rightarrow \widetilde{b}_{k}\left( 0\right) .  \label{mc18}
\end{equation}%
Denoting by $\widetilde{w}_{k}\left( \widetilde{\xi },\tau ^{\ast }\right) $
the result of sunstitutions (\ref{mc18}) in formula (\ref{mc17}) we arrive
at the zeroth-order right corner approximation 
\begin{equation}
\widetilde{u}_{0}^{\left( 2\right) }\left( \rho ,\widetilde{\xi },\tau
^{\ast }\right) =\sum\limits_{k=1}^{\infty }\widetilde{w}_{k}\left( 
\widetilde{\xi },\tau ^{\ast }\right) e^{-\lambda _{k}^{2}\tau ^{\ast }}%
\widehat{J}_{0}\left( \lambda _{k}\frac{\rho }{R_{1}}\right) .  \label{mc19}
\end{equation}%
Note in passing that the limit as $\widetilde{\xi }\rightarrow \infty $
means that $\epsilon \rightarrow 0$ provided $\xi $ is fixed.

\section{Leading-term approximation}

\subsection{Explicit form of the general asymptotic solution}

Inserting partial expansions (\ref{t11}), (\ref{bc6}), (\ref{mc7}) and (\ref%
{ic5}) into general formula (\ref{sd6}) we have uniformly valid as $\epsilon
\rightarrow 0$ in the whole space-time domain $\Sigma _{\tau }$ asymptotic
solution%
\[
u\left( \rho ,\xi ,\tau ;\epsilon \right) =\sum\limits_{m=0}^{\infty
}\left\{ \left[ u_{m}^{\left( 0\right) }\left( \rho ,\xi ,\tau \right)
+u_{m}^{\left( 3\right) }\left( \rho ,\tau ^{\ast };\xi \right) \right]
\epsilon ^{m}\right. 
\]%
\[
+\left[ u_{m}^{\left( 1\right) }\left( \rho ,\xi ^{\ast };\tau \right)
\right. +\widetilde{u}_{m}^{\left( 1\right) }\left( \rho ,\widetilde{\xi }%
;\tau \right) 
\]%
\begin{equation}
+\left. u_{m}^{\left( 2\right) }\left( \rho ,\xi ^{\ast },\tau ^{\ast
}\right) \left. +\widetilde{u}_{m}^{\left( 2\right) }\left( \rho ,\widetilde{%
\xi },\tau ^{\ast }\right) \right] \epsilon ^{m/2}\right\} .  \label{uva1}
\end{equation}%
Finally, collecting here all leading terms, and denoting by $\mathbf{q}%
:=\left\{ \rho ,\xi ,\xi ^{\ast },\widetilde{\xi },\tau ,\tau ^{\ast
}\right\} $ the complete set of outer and inner variables inherent in the
problem under consideration, we can rewrite expansion (\ref{uva1}) in a
compact form 
\begin{equation}
u\left( \rho ,\xi ,\tau ;\epsilon \right) =u_{a}\left( \mathbf{q}\right) +%
\mathcal{O}\left( \sqrt{\epsilon }\right) \text{ }\quad \text{as }\epsilon
\rightarrow 0,  \label{uva2}
\end{equation}%
where $u_{a}\left( \mathbf{q}\right) $ is the leading-term approximation
uniformly valid in domain $\Sigma _{\tau }$ to order $\mathcal{O}\left(
1\right) $. Accordingly, using the obtained results, function $u_{a}\left( 
\mathbf{q}\right) $ may be given as follows: 
\[
u_{a}\left( \mathbf{q}\right) =u_{0}^{\left( 0\right) }\left( \xi ,\tau
\right) 
\]%
\[
+\sum\limits_{k=1}^{\infty }a_{k}\left( \xi \right) e^{-\lambda _{k}^{2}\tau
^{\ast }}\widehat{J}_{0}\left( \lambda _{k}\frac{\rho }{R\left( \xi \right) }%
\right) 
\]%
\[
+\sum\limits_{k=1}^{\infty }\left[ b_{k}\left( \tau \right) e^{-\lambda
_{k}\xi ^{\ast }}+w_{k}\left( \xi ^{\ast },\tau ^{\ast }\right) e^{-\lambda
_{k}^{2}\tau ^{\ast }}\right] \widehat{J}_{0}\left( \lambda _{k}\frac{\rho }{%
R_{0}}\right) 
\]%
\begin{equation}
+\sum\limits_{k=1}^{\infty }\left[ \widetilde{b}_{k}\left( \tau \right)
e^{-\lambda _{k}\widetilde{\xi }}+\widetilde{w}_{k}\left( \widetilde{\xi }%
,\tau ^{\ast }\right) e^{-\lambda _{k}^{2}\tau ^{\ast }}\right] \widehat{J}%
_{0}\left( \lambda _{k}\frac{\rho }{R_{1}}\right) .  \label{uva3}
\end{equation}%
This formula constitutes the main result of the present paper. As an
important consequence of formula (\ref{uva3}) we infer that within the
leading-term approximation the total flux through a tube cross section is
entirely determined by the FJA $u_{0}^{\left( 0\right) }\left( \xi ,\tau
\right) $ and initial boundary layer function $u_{0}^{\left( 3\right)
}\left( \rho ,\tau ^{\ast };\xi \right) $.

Combining expressions (\ref{bc11}), (\ref{bc14}), (\ref{ic11}) and utilizing
the projector $\mathcal{P}_{\xi }$ defined by formula (\ref{Hs7}) one can
see that the FJA $u_{0}^{\left( 0\right) }\left( \xi ,\tau \right) $ is
uniquely determined by the following 1D boundary value problem 
\begin{equation}
\mathcal{L}_{FJ}u_{0}^{\left( 0\right) }=0,  \label{uva4}
\end{equation}%
\begin{equation}
\left. u_{0}^{\left( 0\right) }\right\vert _{\tau =0}=\mathcal{P}_{\xi
}g_{0},  \label{uva4a}
\end{equation}%
\begin{equation}
\left. u_{0}^{\left( 0\right) }\right\vert _{\xi =0}=\mathcal{P}%
_{0}g_{1},\quad \left. u_{0}^{\left( 0\right) }\left( \xi ,\tau \right)
\right\vert _{\xi =1}=\mathcal{P}_{1}g_{2}.  \label{uva4b}
\end{equation}%
Hence one can see that function $u_{0}^{\left( 0\right) }\left( \xi ,\tau
\right) $ is the zeroth-order in $\epsilon $ projection of 3D concentration
on the unit zero eigenfunction of the unperturbed operator $\mathcal{L}%
_{\rho }$ (see Appendix), that is 
\begin{equation}
u_{0}^{\left( 0\right) }\left( \xi ,\tau \right) =\lim_{\epsilon \rightarrow
0}\mathcal{P}_{\xi }u\left( \rho ,\xi ,\tau ;\epsilon \right) .  \label{uva5}
\end{equation}%
It is important to note here that we cannot obtain the FJE (\ref{uva4}) just
by simple projection of the original 3D equation (\ref{nd4}) on the zero
eigenfunction of the unperturbed operator $\mathcal{L}_{\rho }$ because
operators of differentiation with respect to $\xi $ and projection operator
(depending on $\xi $) do not commute.\cite{Kalnay:2005}

Thus, provided one has solved problem (\ref{uva4})-(\ref{uva4b}) with
respect to the FJA $u_{0}^{\left( 0\right) }\left( \xi ,\tau \right) $ the
desired leading-term approximation $u_{a}\left( \mathbf{q}\right) $ is
governed explicitly by formula (\ref{uva3}).

Note in passing that according to expansion (\ref{uva1}) the contribution
from inner solutions in the spatial boundary layer ($u_{m}^{\left( 1\right)
} $ and $\widetilde{u}_{m}^{\left( 1\right) }$) is much more important than
that from the solutions for the initial layer ($u_{m}^{\left( 3\right) }$),
since the influence of the spatial boundary layer at $m=1$ gives a term of
order $\mathcal{O}\left( \sqrt{\epsilon }\right) $.

\subsection{Validity of the Fick-Jacobs approximation}

Let us delineate the conditions on temporal and spatial scales under which
the FJA $u_{0}^{\left( 0\right) }\left( \xi ,\tau \right) $ is valid. One
can see that the general condition for validity of the FJA reads%
\begin{equation}
\left\vert u_{a}-u_{0}^{\left( 0\right) }\right\vert /u_{0}^{\left( 0\right)
}\ll 1.  \label{uva6}
\end{equation}%
To obtain simple validity conditions first we observe that contribution of
the corner functions $u_{0}^{\left( 2\right) }$ and $\widetilde{u}%
_{0}^{\left( 2\right) }$ to $u$ is certainly less than that from the
functions of the diffusion spatial $u_{0}^{\left( 1\right) }$, $\widetilde{u}%
_{0}^{\left( 1\right) }$ and temporal $u_{0}^{\left( 3\right) }$ boundary
layers. Therefore one can ignore in (\ref{uva2}) the corrections due to
corner functions.

It is also clear from (\ref{uva2}) and (\ref{uva3}) that solution $%
u_{0}^{\left( 3\right) }$ corresponds to initial stage of the concentration
evolution in $\Omega ^{\left( 3,0\right) }:=\Omega _{0}^{\left( 3\right)
}\cup \Omega ^{\left( 0\right) }$ (see Fig.~\ref{fig:dom}), where there is a relaxation
to the equilibrium with respect to the transversal variable $\rho $, that is%
\[
\frac{\partial u_{a}\left( \mathbf{q}\right) }{\partial \rho }=0\quad \text{
in }\Omega ^{\left( 3,0\right) }. 
\]%
This process occurs by exponential damping law with times spectrum $%
t_{k}=t_{tr}\lambda _{k}^{-2}$, ($k\geq 1$) and the characteristic
relaxation longitudinal time for homogenization of initially nonuniform in $%
r $ distribution of concentration is determined by the lowest eigenvalue 
\begin{equation}
t_{1}=\frac{t_{tr}}{\lambda _{1}^{2}}\approx 0.0681\cdot \frac{r_{M}^{2}}{%
D_{\perp }}\ll t_{tr}.  \label{uva7}
\end{equation}

Similarly it follows from expression (\ref{uva2}) and form of solutions $%
u_{0}^{\left( 1\right) }$, $\widetilde{u}_{0}^{\left( 1\right) }$ (\ref{uva3}%
) that characteristic thickness of the spatial diffusion boundary layers $%
l_{1}$ along $z$ axis in $\Omega _{0}^{\left( 1\right) }\cup \Omega ^{\left(
0\right) }$ and $\Omega _{1}^{\left( 1\right) }\cup \Omega ^{\left( 0\right)
}$ (see Fig.~\ref{fig:dom}) is 
\begin{equation}
l_{1}=\frac{r_{M}}{\lambda _{1}}\approx 0.2610\cdot r_{M}<r_{M}.
\label{uva8}
\end{equation}%
One can see from (\ref{uva7}) and (\ref{uva8}) that there is a simple
connection between values $t_{1}$ and $l_{1}$: $t_{1}=l_{1}^{2}/D_{\parallel
}$.

Accordingly, combining (\ref{uva7}) and (\ref{uva8}) the validity of the FJA
is determined by the following temporal and spatial conditions 
\begin{equation}
t\gtrsim t_{tr}\gg t_{1},\qquad \ z\text{ (or }L-z\text{)}\gg r_{M}>l_{1}.
\label{uva9}
\end{equation}%
This means that for temporal and spatial scales (\ref{uva9}) the explicit
dependence of the leading-term approximation $u_{a}\left( \mathbf{q}\right) $
on the initial distribution (\ref{sp5}) and boundary conditions (\ref{sp6})
depending on the transversal coordinate $r$ disappeared. In other words the
FJA works well under a quasi steady-state regime with respect to the
characteristic transversal time $t_{tr}$, i.e., when we can eliminate
dependence on fast transversal variable $\rho $ and consider dependence only
upon slow "hydrodynamic" variables $\xi $ and $\tau $.

Particularly, for the cylindric tube of radius $r_{M}$ all deviations from
the "equilibrium function" $v_{0}$ (see Appendix) caused by the inital and
boundary conditions depending on transversal coordinate $\rho $ are vanished
within diffusion boundary layer subdomains. Thus the subdomain $\Omega
^{\left( 0\right) }$ becomes "the equilibrium region", where solution $%
u^{\left( 0\right) }$ does not contain the dependence on $\rho $. It is
clear that this situation occurs due to the wall condition $\left. \left(
\partial u/\partial \rho \right) \right\vert _{\rho =1}=0$.

For general case it is clear that at least in a vicinity near the wall $%
\partial u/\partial \xi \neq 0$ in $\Sigma _{\tau }$ hence and from the
reflection boundary condition (\ref{nd8}) we have%
\[
\left. \frac{\partial u}{\partial \rho }\right\vert _{\rho =R\left( \xi
\right) }\propto R^{^{\prime }}\left( \xi \right) . 
\]%
Therefore $\partial u/\partial \rho \neq 0$\ in a vicinity near the wall and
a deviation of $u\left( \rho .\xi .\tau \right) $ from the eqiliblium value
increases with increasing of function $\left\vert R^{^{\prime }}\left( \xi
\right) \right\vert $. Nevertheless condition $\left\vert R^{^{\prime
}}\left( \xi \right) \right\vert \ll 1$ is not important for validity of the
leading-term approximation $u_{a}\left( \mathbf{q}\right) $.

To describe similar heat transfer problem in the semi-bounded cylinder
Luikov wrote in his book \cite{Luikov:1967}: \textquotedblright Since there
is no loss of heat through the wall of the rod we can treat it as a solid,
where heat spreads only in one direction\textquotedblright\ (see p. 182 of
Ref. 39 ). Then in this book he considered only 1D equation. It infers from
our study that this statement correct only out of the corresponding spatial
and temporal boundary layers, that is in the outer sundomain $\Omega
^{\left( 0\right) }$.

Thus, conditions (\ref{uva9}) determine the temporal and spatial scales when
the FJA holds true, that is 
\[
u_{a}\left( \mathbf{q}\right) \approx u_{0}^{\left( 0\right) }\left( \xi
,\tau \right) .
\]

\subsection{Analogy to the gas kinetic theory}

Previously, using an analogy with the gas kinetic theory, we proposed a
general kinetics equation to describe the kinetics of diffusion-controlled
reactions in case of infinite system for all spatial and temporal scales and
interpreted some results on diffusive interaction in dense arrays of
absorbing particles.\cite{Traytak:1995DI,Traytak:1996} It is appropriate to
note that idea of{} the projection method suggested by Kalnay and Percus was
inspired by analogy with kinetic theory as well. Concerning their method in
Ref. 17 they claimed the following: "It reminds one of Bogolubov's
derivation of the generalized Boltzmann equation, expressing the $n$%
-particle densities as a functional of the one-particle one; here we reduce
similarly the number of coordinates."

For the problem under study we also revealed a profound analogy with the gas
kinetic theory at low Knudsen numbers, that helped us to choose an adequate
mathematical method to find the desired asymptotic solution. The analogy, we
intend to establish, becomes even more profound in the isotropic diffusion
case, i.e., when $D_{\parallel }=D_{\perp }=D$. Hence we consider this case
and, moreover, for the sake of simplicity, here we dwell on 1D gas system
only. Denoting by $T$ a typical time for the gas system of a typical length $%
L$, $w$ a typical molecular velocity, $\lambda $ the mean free path, and $%
t_{\lambda }$ the mean free time we have 
\begin{equation}
T=L/w,\quad t_{\lambda }=\lambda /w.  \label{ab1}
\end{equation}%
Using these scales one can put down linearized Boltzmann's equation with
respect to the distribution function $f\left( \upsilon ,\xi ,\tau \right) $
in the dimensionless form\cite{Cercignani:1969} 
\begin{equation}
\mathcal{Q}_{\upsilon }f+\text{Kn}\left( \frac{\partial f}{\partial \tau }+%
\frac{\partial }{\partial \xi }\upsilon f\right) =0,  \label{ab2}
\end{equation}%
where $\upsilon =v/w$ being the dimensionless velocity, $\tau =t/T$ is the
dimensionless time, $-\mathcal{Q}_{\upsilon }$ is the linearized collision
operator and the small parameter Kn is so-called Knudsen number 
\begin{equation}
\text{Kn}=\frac{t_{\lambda }}{T}=\frac{\lambda }{L}\ll 1.  \label{ab3}
\end{equation}

The analogy between problem (\ref{nd4})-(\ref{nd9}) and the relevant problem
for Eq. (\ref{ab2}) appeared to be striking. Simple comparison of Knudsen
number (\ref{ab3}) with relaxation parameter (\ref{nd11}) shows that the
mean free time $t_{\lambda }$ corresponds to the characteristic transversal
time $t_{tr}$. The value $w_{tr}=D/r_{M}$ may be treated as a typical
transversal diffusion velocity and, therefore, $r_{M}$ corresponds to the
mean free path $\lambda $. In both cases the perturbation operators describe
the particles transport (with the relevant local fluxes for particles
diffusion $-\left( \partial /\partial \xi \right) f$ and for particles flow $%
\upsilon f$) and unperturbed operators $\mathcal{L}_{\rho }$ and $-\mathcal{Q%
}_{\upsilon }$ are in spectrum. Namely for $-\mathcal{Q}_{\upsilon }$ number 
$\lambda _{0}=0$ is a degenerate eigenvalue with five associated
eigenfunctions, meanwhile for $\mathcal{L}_{\rho }$ there is only one
eigenfunction associated with one trivial eigenvalue. That is why in both
cases for the zeroth-order outer approximation we can derive only equations
for the corresponding projections on the above eigenfunctions (see (\ref%
{uva5})). Moreover, one can see that so-called Hilbert asymptotic solution
(ideal liquid approximation) of Eq. (\ref{ab2}) entirely corresponds to the
FJA $u_{0}^{\left( 0\right) }$ and the normal region\cite{Cercignani:1969}
is nothing more than $\Omega ^{\left( 0\right) }$. In the kinetic theory as
in the problem under study these terms, however, cannot describe the
solutions into initial and boundary layers which naturally arise in both
cases as well.\cite{Cercignani:1969}

It is well known that the classical Chapman-Enskog method is widely used to
reduce the Boltzmann kinetic equation to appropriate hydrodynamic and
transport equations. Noteworthy that if we consider higher-order
approximations of the Chapman-Enskog method, we obtain differential
equations of higher order. Nevertheless, it is long known that the
Chapman-Enskog expansion can bring in solutions, which are nonexistent. In
order to overcome these difficulties the method of matching inner and outer
expansions was also applied in kinetic theory.\cite{Cercignani:1969} It
seems that the Kalnay-Percus mapping approach\cite%
{Kalnay:2005,Kalnay2:2005,Kalnay:2006} resembles some features of the
Chapman-Enskog method, but this question needs to be investigated.

However, the analogy at issue is limited. The reflecting boundary condition (%
\ref{nd8}) plays an essential role in the diffusion problem (\ref{nd4})-(\ref%
{nd9}). Exactly due to this condition in contrast to the unperturbed
operator of kinetics theory $-\mathcal{Q}_{\upsilon }$, operator $\mathcal{L}%
_{\rho }$ (\ref{nd5}) is not self-adjoint (see Appendix). Nevertheless,
detected analogy enables us to elucidate a number of the features inherent
in the asymptotic solution of the diffusion problem (\ref{nd4})-(\ref{nd9})
at small values of the relaxation parameter $\epsilon $.

\section{Concluding remarks}

By means of matched asymptotic expansions approach we gained here the
uniformly valid leading-term approximation (\ref{uva3}) to solution of the
3D diffusion problem (\ref{nd4})-(\ref{nd9}) with respect to small
relaxation parameter of the tube $\epsilon $ (\ref{nd11}).

Suggested here derivation elucidates the mathematical sense of the
Fick-Jacobs equation as the solvability condition of the lowest order (\ref%
{t15}) for the first correction in $\epsilon $ to the Fick-Jacobs
approximation. Asymptotic solution also shows that known quasi-cylindrical
condition $\left\vert R^{^{\prime }}\left( \xi \right) \right\vert \ll 1$ is
not necessary for validity of the leading-term approximation including
Fick-Jacobs approximation. At the same time matching procedure automatically
gave us an exact algorithm for determination of the missing initial and
boundary conditions which must be imposed on the Fick-Jacobs approximation
and its corrections.

The explicit form of the leading-term approximation allowed us to delineate
the conditions on temporal and spatial scales (\ref{uva9}) under which the
Fick-Jacobs approximation is valid.

One of the most noteworthy features of all previously suggested zeroth-order
corrections to the Fick-Jacobs approximation is the absence of dependence on
the transverse coordinates. However, as we have shown, even the leading-term
approximation comprises the boundary layers solutions explicitly depending
on the transversal coordinate. We also proved that the outer approximation $%
u^{\left( 0\right) }\left( \rho ,\xi ,\tau \right) $ starting from orders $%
\mathcal{O}\left( \epsilon \right) $ contains terms explicitly depending on
the transversal variable $\rho $ (see (\ref{t16})).

A profound analogy between the problem under consideration and the method of
inner-outer expansions for low Knudsen numbers gas kinetic theory is
established. This analogy enables us to clarify the physical and
mathematical meaning of the obtained results.

It is important to underline that contrary to other known approaches our
derivation of the Fick-Jacobs equation was implemented straightforwardly
within the scope of asymptotic method procedure without any additional
assumptions. In this connection we believe that it is rather inexpedient to
exploit any physical arguments during the solution of quite well posed
mathematical problem.

Future extension of the present work may include the higher order in $%
\epsilon $ corrections to the solution considered here and also the case of
tubes of other varying constraint geometry, e.g., without axial symmetry.
The results obtained in this paper allow us to hope that the matched
asymptotic expansions method may be successfully applied to many other
problems concerning diffusion transport of pointlike particles in tubes of
varing cross section. For example, the diffusion of particles undergoing the
influence of interaction potential, partially penetrable boundary condition
on the tube wall and its ends or diffusion equation with a source term may
be considered by means of above method.

\section*{ACKNOWLEDGMENTS}

This research has been partially supported by \textquotedblright Le
STUDIUM\textquotedblright\ (Loire Valley Institute for Advanced Studies). We
also personally thank Professors P. Vigny and N. Fazzalari for their 
interest in this study and Professor F. Piazza for useful discussions.

\section*{APPENDIX}
\appendix*

For the sake of completeness we recall here some useful classical
mathematical definitions and facts.

The boundary layer of a domain $\Sigma _{\tau }$ comprises the set of points
from $\Sigma _{\tau }$ such that their distance to the boundary $\partial
\Sigma _{\tau }$ does not exceed some given magnitude $\delta >0$, which is
called the thickness of the layer.

In theory of singular perturbed problems a function $u_{a}\left( \mathbf{x}%
;\epsilon \right) $ is said to be an approximation to $u\left( \mathbf{x}%
;\epsilon \right) $ uniformly valid in a domain $\Lambda \subset \mathbb{R}%
^{n}$ to order $\mathcal{O}\left( \zeta \left( \epsilon \right) \right) $ as 
$\epsilon \rightarrow 0$ if 
\begin{equation}
\lim_{\epsilon \rightarrow 0}\frac{\left\vert u\left( \mathbf{x};\epsilon
\right) -u_{a}\left( \mathbf{x};\epsilon \right) \right\vert }{\zeta \left(
\epsilon \right) }=0  \label{con1}
\end{equation}%
uniformly for all $\mathbf{x}\in \Lambda $.\cite{Lagerstrom:1988} Here $%
\zeta \left( \epsilon \right) $ is so-called a gauge function.

Let us introduce the space $\mathcal{H}_{\xi }$ of twice continuously
differentiable real-valued functions $v:\left( 0,R\left( \xi \right) \right)
\rightarrow \mathbb{R}_{+}$ given on the cross section of the tube $0<\rho
<R\left( \xi \right) $ at any fixed point $\left( \xi ,\tau \right) $.
Additionally we assume that functions $v\in \mathcal{H}_{\xi }$ obey the
Neumann boundary conditions 
\begin{equation}
\left. v\right\vert _{\rho =0}<\infty ,\qquad \left. \frac{\partial v}{%
\partial \rho }\right\vert _{\rho =0}=0,  \label{Hs0a}
\end{equation}%
\begin{equation}
\left. \frac{\partial v}{\partial \rho }\right\vert _{\rho =R\left( \xi
\right) }=0  \label{Hs0b}
\end{equation}%
at fixed point $\left( \xi ,\tau \right) $.

Then for any two functions $f$, $g\in \mathcal{H}_{\xi }$ we can introduce
the weighted $L_{\rho }^{2}$ scaler product with the weight function $\rho $
as 
\begin{equation}
\left\langle f,g\right\rangle _{\mathcal{H}_{\xi
}}:=\int\limits_{0}^{R\left( \xi \right) }\rho f\left( \rho \right) g\left(
\rho \right) d\rho .  \label{Hs1}
\end{equation}%
One can show that this defines the weighted Hilbert space $\mathcal{H}_{\xi
}:=L_{\rho }^{2}\left( \left( 0,R\left( \xi \right) \right) \right) $ with
the norm \cite{Rektorys:1977} 
\[
\left\Vert f\right\Vert _{\mathcal{H}_{\xi }}=\left[ \int\limits_{0}^{R%
\left( \xi \right) }\rho f^{2}\left( \rho \right) d\rho \right]
^{1/2}<\infty . 
\]

Consider the linear operator $\mathcal{L}_{\rho }:\mathcal{H}_{\xi
}\rightarrow C\left( 0,R\left( \xi \right) \right) $ defined by (\ref{nd5}).
One can see that operator $\mathcal{L}_{\rho }$ is self-adjoint in $\mathcal{%
H}_{\xi }$, that is%
\begin{equation}
\left\langle \mathcal{L}_{\rho }f,g\right\rangle _{\mathcal{H}_{\xi
}}=\left\langle f,\mathcal{L}_{\rho }g\right\rangle _{\mathcal{H}_{\xi }}.
\label{Hs2a}
\end{equation}%
Hence there exists the nontrivial solution of the eigenvalue problem 
\begin{equation}
\mathcal{L}_{\rho }v=\lambda ^{2}v  \label{Hs3}
\end{equation}%
for $v\in \mathcal{H}_{\xi }$ under the Neumann boundary conditions (\ref%
{Hs0a}) and (\ref{Hs0b}).

It has real pure-point spectrum of eigenvalues $\left\{ \lambda _{k}\right\}
_{k=0}^{\infty }$ such that for all $k\geq 0$ we have the ordering 
\[
0\leq \lambda _{0}<\lambda _{1}<...<\lambda _{k}<... 
\]%
at that $\lambda _{k}\rightarrow \infty $ as $k\rightarrow \infty $.

The associated eigenfunctions $v_{k}\in \mathcal{H}_{\xi }$ of the problem (%
\ref{Hs3}) are 
\begin{equation}
v_{k}:=J_{0}\left( \lambda _{k}\frac{\rho }{R\left( \xi \right) }\right) .
\label{Hs3a}
\end{equation}%
Here $J_{\nu }\left( \zeta \right) $ is Bessel's function of the first kind
of order $\nu $ which may be defined by its Maclaurin series\cite%
{Arfken:2001}%
\begin{equation}
J_{\nu }\left( \zeta \right) =\sum\limits_{m=0}^{\infty }\frac{\left(
-1\right) ^{m}}{\Gamma \left( m+\nu +1\right) m!}\left( \frac{\zeta }{2}%
\right) ^{2m+\nu },  \label{Hs3b}
\end{equation}%
where $\Gamma \left( \beta \right) $ is the gamma function.

Thus the eigenvalues $\lambda _{k}$ of the eigenvalue problem (\ref{Hs3})
are determined by the transcendental equation 
\begin{equation}
J_{0}^{^{\prime }}\left( \lambda _{k}\right) =0  \label{Hs4}
\end{equation}%
which follows from the Neumann condition (\ref{Hs0b}). Taking advantage of
the known relation $J_{0}^{^{\prime }}\left( \zeta \right) =-J_{1}\left(
\zeta \right) $ we infer that $\lambda _{k}$ are also the roots of the
transcendental equation 
\begin{equation}
J_{1}\left( \lambda _{k}\right) =0.  \label{Hs4a}
\end{equation}%
It follows from expansion (\ref{Hs3b}) and Eq. (\ref{Hs4a}) that $\lambda
_{0}=0$ and $v_{0}=J_{0}\left( 0\right) =1$. One can see that for $\lambda
_{0}=0$ there is only one linear independent eigenfunction $v_{0}\equiv const
$.

We have and other eigenvalues , e.g. \cite{Abram:1972} 
\[
\lambda _{1}\approx 3.8317,\text{ }\lambda _{2}\approx 7.0156,\text{ }%
\lambda _{3}\approx 10.1735,...\text{ } 
\]%
Eigenfunctions $\left\{ v_{k}\right\} _{k=0}^{\infty }$ form a complete
orthogonal system in $\mathcal{H}_{\xi }$ and the orthogonality property for
them holds\cite{Arfken:2001} 
\begin{equation}
\left\langle v_{k},v_{m}\right\rangle _{\mathcal{H}_{\xi }}=\delta
_{km}\left\Vert v_{k}\right\Vert _{\mathcal{H}_{\xi }}^{2},  \label{Hs5}
\end{equation}%
where $\delta _{km}$ is the Kronecker delta and 
\begin{equation}
\left\Vert v_{k}\right\Vert _{\mathcal{H}_{\xi }}^{2}=\frac{1}{2}R^{2}\left(
\xi \right) J_{0}^{2}\left( \lambda _{k}\right) .  \label{Hs5a}
\end{equation}%
The corresponding orthonormal system $\left\{ \widehat{v}_{k}\right\}
_{k=0}^{\infty }$ is defined by 
\begin{equation}
\widehat{v}_{k}=v_{k}/\left\Vert v_{k}\right\Vert _{\mathcal{H}_{\xi }}.
\label{Hs6}
\end{equation}

For any $\lambda _{k}$ there is only one normed eigenfunction $\widehat{v}%
_{k}\in \mathcal{H}_{\xi }$ therefore for any function $f\in L_{\rho
}^{2}\left( \left( 0,R\left( \xi \right) \right) \right) $ we have the
Fourier series with respect to orthonormal basis $\left\{ \widehat{v}%
_{k}\right\} _{k=0}^{\infty }$, that is%
\begin{equation}
f=\sum\limits_{k=0}^{\infty }c_{k}\widehat{v}_{k},  \label{Hs6a}
\end{equation}%
where $c_{k}=\left\langle f,\widehat{v}_{k}\right\rangle $. It is known that
series (\ref{Hs6a}) converges in $\mathcal{H}_{\xi }$, so the set of
orthonormal functions $\left\{ \widehat{v}_{k}\right\} _{k=0}^{\infty }$ is
complete.\cite{Rektorys:1977}

Note that existence of the trivial eigenvalue $\left\{ \lambda
_{0}=0\right\} $ for (\ref{Hs3}) leads to a serious complication for
solution of Eq. (\ref{nd4}). Operator $\mathcal{L}_{\rho }$ is termed to be
in spectrum if there is at least one trivial eigenvalue belongning to the
spectrum of $\mathcal{L}_{\rho }$.\cite{Vishik:1960}

Consider a linear subspace $V$ of the Hilbert space $\mathcal{H}_{\xi }$
spanned on the zero eigenfunction $v_{0}$, that is $V=\left\{ v\in
V:v=\alpha v_{0},\alpha \in \mathbb{R}\right\} $. It is well known that
there exists space $V^{\bot }$ orthogonal to $V$ such that $\left( V^{\bot
}\right) ^{\perp }=V$ and $\mathcal{H}_{\xi }=V\oplus V^{\perp }$. So we can
define the linear orthogonal projection operator (projector) $\mathcal{P}%
_{\xi }:\mathcal{H}_{\xi }\rightarrow V$ that maps any $w\in \mathcal{H}%
_{\xi }$ to $v\in V$ is called the orthogonal projection onto $V$, i.e.%
\begin{equation}
\mathcal{P}_{\xi }w=\left\langle w,v\right\rangle _{\mathcal{H}_{\xi }}v.
\label{Hs6b}
\end{equation}%
In this paper it is convenient to define the projector as follows: 
\begin{equation}
\mathcal{P}_{\xi }w=\left\langle w,\widehat{v}_{0}\right\rangle _{\mathcal{H}%
_{\xi }}\widehat{v}_{0},  \label{Hs7}
\end{equation}%
where $\widehat{v}_{0}=\widehat{J}_{0}\left( 0\right) =\sqrt{2}/R\left( \xi
\right) $.

We observe here that simple Neumann condition (\ref{Hs0b}) arises in the
simplified problems corresponding to diffusion boundary layers. In general
case of functions $f$ and $g$ obeying the reflecting boundary condition (\ref%
{nd8}) the unperturbed operator $\mathcal{L}_{\rho }$ (\ref{nd5}) is not
self-adjoint, i.e. 
\begin{equation}
\left\langle \mathcal{L}_{\rho }f,g\right\rangle _{\mathcal{H}_{\xi }}\neq
\left\langle f,\mathcal{L}_{\rho }g\right\rangle _{\mathcal{H}_{\xi }}.
\label{Hs2}
\end{equation}%
The latter property of the unperturbed operator $\mathcal{L}_{\rho }$
greatly complicates the original boundary value problem (\ref{nd4})-(\ref%
{nd9}).

\end{document}